\documentclass[twocolumn]{aastex62}
\pdfoutput=1
\journalinfo{\@accepted}
\newcommand{\ha}{\ensuremath{\textrm{H}\alpha}}

\newcommand{\kms}{\ensuremath{\textrm{km s}^{-1}}}
\newcommand{\dg}{\ifmmode{^{\circ}}\else $^{\circ}$\fi}
\newcommand{\msun}{\ensuremath{\textrm{M}_\sun}}

\newcommand{\hi}{\ion{H}{1}}
\newcommand{\hii}{\ion{H}{2}}
\newcommand{\lb}{\ifmmode{(\ell,b)} \else $(\ell,b)$\fi}

\newcommand{\cii}{\ion{C}{2*}}

\newcommand{\vlsr}{\ifmmode{v_{\rm{LSR}}}\else $v_{\rm{LSR}}$\fi}
\newcommand{\av}{\ifmmode{A(V)}\else $A(V)$\fi}
\newcommand{\ebv}{\ifmmode{E(B-V)}\else $E(B-V)$\fi}
\newcommand{\iha}{\ensuremath{I_{\mathrm{H}\alpha}}}
\newcommand{\nhi}{\ensuremath{N_{\rm{H\,\sc{I}}}}}

\newcommand{\ihb}{\ifmmode{I_{\rm{H}\beta}} \else $I_{\rm H \beta}$\fi}
\newcommand{\isii}{\ifmmode{I_{\ion{\rm{S}}{2}}} \else $I_{\rm [S \textsc{ ii}]}$\fi}
\newcommand{\inii}{\ifmmode{I_{\ion{\rm{n}}{2}}} \else $I_{\rm [N \textsc{ ii}]}$\fi}
\newcommand{\ioi}{\ifmmode{I_{\ion{\rm{o}}{1}}} \else $I_{\rm [O \textsc{ i}]}$\fi}

\defcitealias{barger2013warmionized}{BHB13} 

\submitjournal{ApJ}
\accepted{October 5, 2019}

\shorttitle{The Diffuse Ionized Gas Halo of the Small Magellanic Cloud} 
\shortauthors{Smart et al.}

\begin{document}

\title{The Diffuse Ionized Gas Halo of the Small Magellanic Cloud}

\correspondingauthor{B. M. Smart}
\email{bsmart@astro.wisc.edu}

\author{B. M. Smart}
\affil{Department of Astronomy, University of Wisconsin-Madison, Madison, WI 53706, USA}
\affil{Department of Physics, Astronomy, and Mathematics, University of Hertfordshire, Hatfield AL10 9AB, UK}

\author{L. M. Haffner}
\affil{Department of Physics and Astronomy, Embry-Riddle Aeronautical University, Daytona Beach, FL 32114-3900, USA}
\affil{Department of Astronomy, University of Wisconsin-Madison, Madison, WI 53706, USA}
\affil{Space Science Institute, Boulder, CO 80301, USA}

\author{K. A. Barger}
\affil{Department of Physics and Astronomy, Texas Christian University, Fort Worth, TX 76129, USA}

\author{A. Hill}
\affil{Department of Physics and Astronomy, University of British Columbia, Vancouver, BC V6T 1Z1, Canada}
\affil{Space Science Institute, Boulder, CO 80301, USA}
\affil{Dominion Radio Astrophysical Observatory, Herzberg Program in Astronomy and Astrophysics, National Research Council Canada, Penticton, BC}

\author{G. Madsen}
\affil{Institute of Astronomy, University of Cambridge, Madingley Road, Cambridge CB3 0HA, UK}

\begin{abstract}

Observations with the Wisconsin \ha\ Mapper (WHAM) reveal a large, diffuse ionized halo 
that surrounds the Small Magellanic Cloud (SMC). We present the first kinematic \ha\  survey of an extended
region around the galaxy, from $(\ell, b) = (289\fdg5, -35\fdg0)$ to 
$(315\fdg1, -5\fdg3)$ and covering
$+90\leq \vlsr \leq +210\ \kms$.
The ionized gas emission extends far beyond the central stellar component
of the galaxy, reaching similar distances to that of the diffuse neutral
halo traced by 21 cm observations. \ha\ emission extends several degrees
 beyond the sensitivity of current \hi\ surveys toward smaller Galactic longitudes and more negative Galactic latitudes.
The velocity field of the ionized gas near the SMC appears similar to
to the neutral halo of the galaxy. 
Using the observed emission measure as a guide, we estimate the mass
of this newly revealed ionized component to be roughly $(0.8-1.0)\times10^{9}\,\msun$,
which is comparable to the total neutral mass in the same region of
$(0.9-1.1)\times10^{9}\,\msun$. We find ratios
of the total ionized gas mass divided by the total neutral plus ionized gas mass
in three distinct subregions to be: (1) 46\%--54\% throughout the SMC
and its extended halo, (2) 12\%--32\% in the SMC Tail that extends
toward the Magellanic Bridge, and (3) 65\%--79\% in a filament that
extends away from the SMC toward the Magellanic Stream. This newly discovered,
coherent \ha\ filament does not appear to have a well-structured
neutral component and is also not coincident with two previously identified
filaments traced by 21 cm emission within the Stream. 
\end{abstract}


\section{Introduction}

\begin{figure}[t]
%\begin{centering}
%\begin{tabular}{c}
\includegraphics[width=\columnwidth,clip]{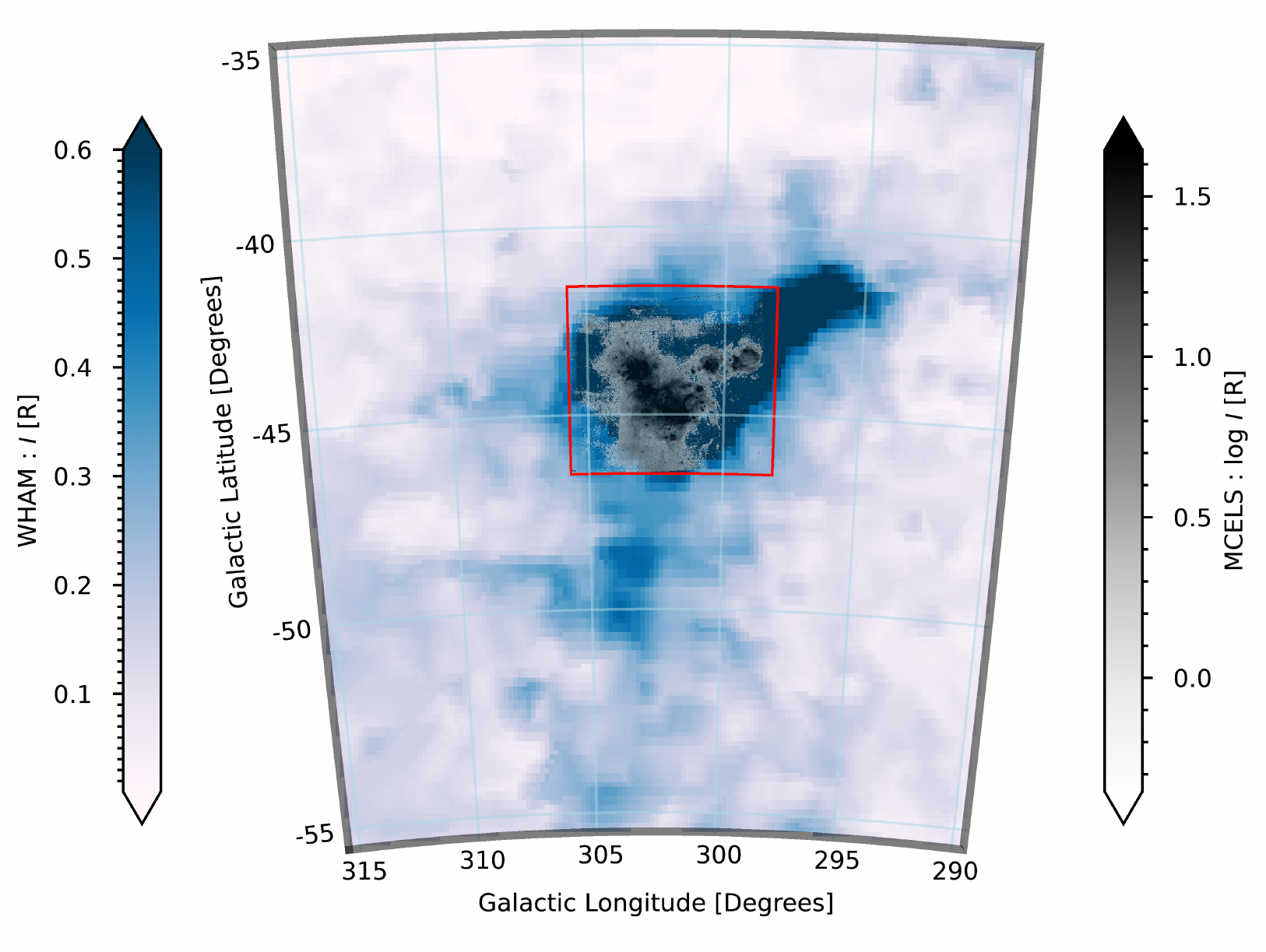}
%\tabularnewline
%\end{tabular}
%\par\end{centering}

\caption{WHAM \ha\  emission (\emph{blue, background}) associated with the
SMC, integrated between $+90\ \kms \leq \vlsr \leq +120\ \kms$
and the MCELS \ha\  image
(\emph{grayscale, foreground}) from \citet{winkler2015theinterstellar}.
The red box denotes the extent of the MCELS survey region. Note that
the WHAM emission scaling is linear (\emph{left colorbar}) while
scaling for the MCELS image is logarithmic (\emph{right colorbar})
to highlight the bright structures within the galaxy. \label{fig:MCELS} }

\end{figure}

The Small Magellanic Cloud (SMC) is a nearby, low-mass irregular galaxy with
a gas-rich interstellar medium (ISM). As a member of an
interacting system that includes the Large Magellanic Cloud (LMC),
Small Magellanic Cloud, and the Milky Way (MW), the SMC has
undergone interactions that have created an extended envelope of
gas \citep{gardiner1994numerical,gardiner1996nbodysimulations,besla2007arethe}.
The extent of the \hi\ envelope has been thoroughly studied
\citep{hindman1963alow,stanimirovic1999thelargescale,stanimirovic2004anew,bruens2005theparkes,kalberla2005theleidenargentinebonn,nidever2010the200textdegree}
with the total neutral gas mass of the galaxy measured to be $4.0\times10^8~{\rm M}_\odot$
\citep{bruens2005theparkes}. However, studies of the ionized portion
of the gas have been primarily limited to H\textsc{~ii} regions,
supernovae remnants, large filamentary structures, \citep{lecoarer1993halphasurvey}  or to the SMC Tail \citep[hereafter BHB13]{barger2013warmionized}.
The Magellanic Cloud Emission-line Survey (MCELS) is the most recent
imaging survey of the SMC in \ha, spanning a $4\fdg5 \times 3\fdg5$\ region
\citep{winkler2015theinterstellar}. Compared to the extent of the
\hi\ envelope, MCELS surveyed the ionized gas in the central region of the galaxy 
and does not trace the diffuse emission from the warm ionized medium (WIM; see Figure \ref{fig:MCELS}).

For many active star-forming galaxies, a considerable amount of total
ionized gas mass is contained in the WIM (conventionally called diffuse ionized gas (DIG) in external galaxies)
and gives us insight into the radiation flowing through the galaxy \citep{Rossa2000,Rossa2003a,Rossa2003b,Zurita2000}.
As is apparent from our own MW's WIM, diffuse ionized gas is pervasive
beyond \hii\ regions and supernovae remnants \citep{haffner2003thewisconsin, Haffner2009,hill2012modernview,Krishnarao2017},
and can extend far beyond the limits of the detectable neutral gas.
In the MW, the WIM has a filling factor of up to $\sim30\%$ \citep{reynolds1977pulsardispersion,Berk1998LNP...506..301B,Gaensler2008,Savage&Wakker2009,Hill2014,Krishnarao2017}
that is dependent on scale height. Emission is visible in every direction
from our location in the Galaxy \citep{haffner2003thewisconsin}.
Its pervasive presence lends insight into the source of ionization
and the structure of the interstellar medium (ISM). However, it is
difficult to get a complete census of the WIM in the MW and identify the sources of its ionization due to our vantage point.
Nearby galaxies, like the SMC, give us a chance to view the global distribution 
of the WIM in a galaxy and its collective sources of ionization from
an outside perspective.

As few studies have explored the extended DIG 
 of low-mass galaxies, it is uncertain how closely
the physical properties and the ionization conditions of their gas
resembles that of the MW. There are a limited number of studies
on the DIG in dwarf irregular galaxies which used aperture photometry
\citep{hunter1990properties,hunter1992supergiant,hunter1993asurvey}
or slit spectroscopy \citep{Martin1997,Domgorgen1997} to measure
\ha\  emission. \citet{hunter1990properties} found that 15\%-20\%
of the \ha\  flux emitted from dwarf irregular galaxies is from the diffuse
ionized medium. Using assumptions of the properties of the DIG, they
suggest that diffuse material is the dominant form of ionized gas
in dwarf irregulars. However, these surveys were primarily limited
to bright structures within the galaxies rather than the faint portion
of the DIG. This restriction to dense structures is due to difficulty
in detecting the DIG in dwarf irregulars. \citet{hunter1993asurvey}
took deep \ha\  images of 51 irregular and amorphous galaxies, detecting
emission down to a few Rayleighs\footnote{1 Rayleigh = $\frac{10^{6}}{4\pi}$ photons s$^{-1}$ cm$^{-2}$ sr$^{-1}$}
for the more nearby galaxies. A more recent study by \citet{Oey2007},
 which defined the DIG as \ha\-emitting regions separate from \hii\ regions,
 looked at a wide range of star forming galaxies and found $59\%\pm19\%$
of the \ha\  originates from the WIM, a higher fraction than previous
studies. Prior studies have generally not been very sensitive to faint DIG.
 However, they have been able to characterize the diffuse filamentary structure far from classical \hii\ regions. 
 Conversely, in our study, we are not able to exclude \hii\ region emission in our observations of the central 
 region of the SMC due to our beam size, which corresponds to a spatial resolution of 1 kpc at the distance of 
 the galaxy. In the halo, this limitation does not impact our results. 
\begin{figure}[t]
%\begin{centering}
%\begin{tabular}{c}
\includegraphics[width=\columnwidth,trim={2cm 7cm 2cm 5cm},clip]{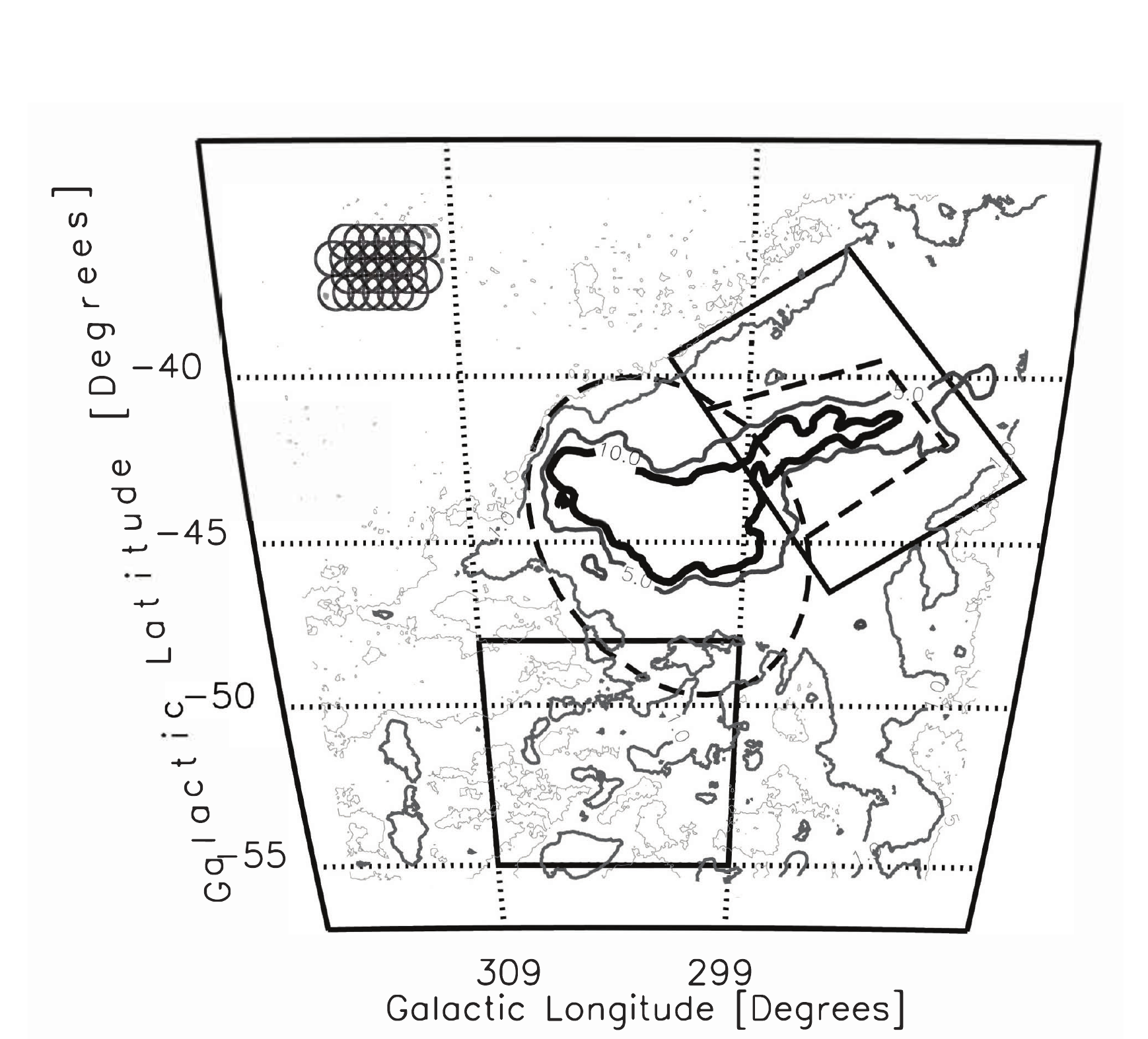}
%\tabularnewline
%\end{tabular}
%\par\end{centering}

\centering{}\caption{Full map of the SMC with each subregion highlighted. The central
SMC region is marked by the dashed ellipse. The upper right solid
line marks the \hi\ SMC Tail and inner dashed lines mark
the \ha\  SMC-Tail. The SMC-Stream interface region at the bottom
is marked in solid black lines. The contours mark the \hi\
column density at 0.5, 1, 5 and $10\times10^{20}\ \mathrm{cm}^{-2}$.
The grid of $1\arcdeg$
circles at the top left of the map represent the Nyquist sampling,
with points separated by $0\fdg5$ beam-steps, which
were used to map the \ha\  emission of WHAM observations.\label{fig:Skematic}}
\end{figure}

In the Cosmic Origins Spectrograph (COS)/ Ultraviolet and Visual Echelle Spectrograph (UVES) Absorption Survey, \citet{fox2014thecosuves} used
absorption toward 69 active galactic nuclei (AGN) to push  push the detection limit of the ionized gas 
associated with the Magellanic Clouds much lower toward regions with \hi\  column densities of $\log \nhi \approx 19.4$--$20.1$.
Their target AGNs are located near the Magellanic System and along 
the Stream and sparsely sample the large structures. Away from the prominent 21 cm 
emitting structures, their lowest neutral column density sightlines (log \nhi\ $ < 19.5$) are found to 
have very high ionization fractions (up to 98\%).

\citetalias{barger2013warmionized} 
reported the first extended detection of DIG near the Magellanic Clouds. Their 
 study of the Magellanic Bridge revealed that DIG traces
the \hi\ 21 cm emission, though they also found that the
\ha\  often extends many several degrees beyond the detectable neutral
hydrogen emission. Additionally, the study suggested an upper limit
of 50\% ionization of the gas within the Bridge system \citepalias{barger2013warmionized}. This
is similar to the general pattern in high-velocity clouds around the
MW, in which the detection of \ha\  emission is generally
correlated with the detection of \hi\ but the intensities are not correlated
\citep{Tufte1998ApJ...504..773T,hill2009ionized,Barger2012ComplexA,barger2017}. If similar conditions
exist within the SMC, then a large fraction of the gas has gone undetected
by previous studies. 

To determine if the DIG throughout the SMC has a similar ionization
fraction to that of the Milky Way and the Magellanic Bridge, we conducted
an \ha\ emission survey of the SMC using WHAM. In Section
\ref{section:Observe}, we describe the WHAM observations and outline our data reduction
process in Section \ref{section:Data}. We present the non-extinction corrected \ha\
intensity map of the SMC in Section \ref{section:Intensity} and discuss the differences
and similarities of the \ha\ and \hi\ emission and velocity distributions.
 We investigate the total mass of the SMC by addressing the distribution
of neutral and ionized gas in Section \ref{sub:Mass}. We discuss our work in Section \ref{section:Discussion} and
 present our conclusions in conclusions in Section \ref{section:Conc}.

\section{Observations \label{section:Observe}}

\begin{figure*}[t]

\includegraphics[trim={0 5cm 0 5.2cm},clip,width=0.8\paperwidth]{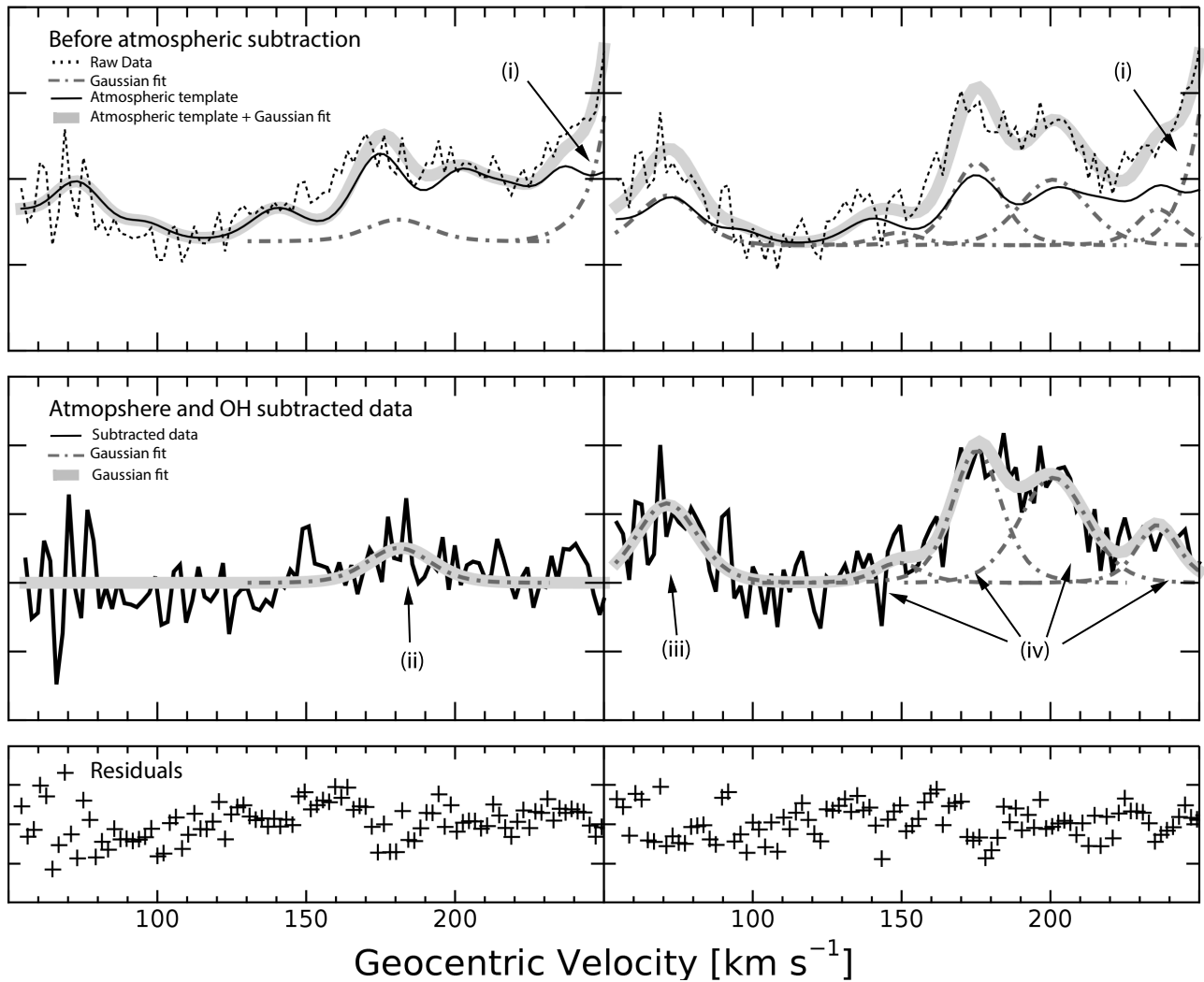}

\caption{Example of atmospheric subtraction from a region with faint emission
and a region with bright emission. The emission in the first column
was taken off the SMC at $(\ell, b) = (298\fdg88, -37\fdg44)$. The
second column was taken in a bright \ha\  emitting direction at $(\ell, b) = (293\fdg30,
-41\fdg11)$. The pre-atmospheric subtracted images are shown in panels the first row. The 
solid black line indicates the atmospheric template (see Section \ref{sub:Faint-Atmospheric-Line-Section}), and the solid gray line
marks the average fit combining the atmospheric template, the OH line fit, and emission fits. We describe our procedure for fitting the sky
 lines in Sections \ref{sub:Bright-Atmospheric-Line-Section} and \ref{sub:Faint-Atmospheric-Line-Section}, with references to \citet{hausen2002interstellar}, \citet{haffner2003thewisconsin}, and \citetalias{barger2013warmionized}, where the procedure is described in detail.
The second row displays the post atmospheric and OH subtracted 
emission, with the dashed Gaussians marking fitted emission.
The residuals of the subtraction are displayed in the third row. The
(i) marker indicates the a bright OH line at a geocentric velocity
of +272.44 \kms\ and (ii) indicates faint
emission. Galactic emission is labeled by markers (iii). Marker (iv)
indicates the bright emission from the SMC.\label{fig:brightfaintcomparison}}
\end{figure*}

The WHAM facility was specifically designed with high sensitivity
to faint optical emission lines from diffuse sources. 
Because the intensity of these emission lines scale with 
 the emission measure ($EM=\int n_{e}^{2}\,dl$),
 the low-density gas in the galactic halos is very faint. An instrument with high sensitivity is thus necessary to observe
them. The previous SMC studies listed in Section 1 detected \ha\  with
lower intensity limit of $\sim$3--6 R, limiting observations to the
central stellar region of the SMC. WHAM can detect features reliably
down to $\sim25$ mR, limited primarily by foreground atmospheric
lines for individual observations. However, for extended, continuous
sources, WHAM has detected structures below 10 mR \citep{Barger2012ComplexA}.

The spectrometer, described in detail by \citet{haffner2003thewisconsin},
consists of a dual-etalon Fabry-Perot spectrometer that produces a
200 \kms\ wide spectrum with 12 \kms\
velocity resolution from light integrated over a 1\arcdeg\ beam.
 Typical \ha\  line widths from diffuse ionized gas have FWHM of $\geq 20$
\kms,  which is well-matched to WHAM's spectral resolution.
A 30-second exposure can achieve a signal to noise of 20 for a 0.5
R line with a width of 20 \kms. 

Observations of the SMC were taken between October 1st, 2010, and
January 5th, 2011 at Cerro Tololo Inter-American Observatory. We Nyquist
sampled the \ha\ emission of the SMC with WHAM at an angular
resolution of 1\arcdeg\  over the local
standard of rest (LSR) velocity range of $+50 \leq \vlsr \leq +250$ \kms\
from $( \ell , b ) = (289.5\arcdeg , -35.0\arcdeg)$ to $(315.1\arcdeg,
-55.3\arcdeg)$. Figure \ref{fig:Skematic}
illustrates the angular extent and spacing of a ``block" of observations
in the top left.

We used the observing strategy for Nyquist sampled beam spacing described
by \citetalias{barger2013warmionized}, grouping observations into ``blocks''
of 30--50 pointings at 0\fdg5
spacings. Each single observation had an exposure time of 30 s. Keeping
exposure times short reduces the variations observed in atmospheric
lines within a single block observation.

\section{Data Reduction \label{section:Data}}

\begin{table*}[t]
%\centering{}
\caption{Extinction\label{tab:Extinction} }
\begin{center}
\begin{tabular}{ccccccc}

\hline 
\hline 
{Region} & \multicolumn{3}{c}{Foreground Extinction} & \multicolumn{3}{c}{Internal Extinction}\tabularnewline
\cline{2-7} 
 & log $\left\langle \nhi \right\rangle $ & $A(\ha)$\tablenotemark{a} & $\%{}_{corr}$ & log $\left\langle \nhi \right\rangle $ & $A(\ha)$  & $\%{}_{corr}$\tabularnewline
\hline
SMC & 20.5 & 0.15-0.14 & 12.9-14.4\% & 21.0 & 0.24-0.27\tablenotemark{b} & 20-23\%\tabularnewline
\hi\ SMC Tail & 20.6 & 0.17-0.20 & 14.6-16.6\% & 20.5 & 0.06-0.03 & 6.0-5.6\%\tabularnewline
\ha\  SMC Tail & 20.6 & 0.17-0.19 & 13.7-15.8\% & 20.7 & 0.11-0.6 & 10.0-9.7\%\tabularnewline
SMC Filament\tablenotemark{c} & 20.2 & 0.07-0.08 & 6.4-7.6 \% & - & - & -\tabularnewline
\hline 
\end{tabular}
\end{center}

\tablenotetext{a}{The first listed value was determined by averaging $\left\langle \nhi \right\rangle $
in the region used to calculate masses in all scenarios other than
the ellipsoidal scenario((See Section \ref{sub:Mass-of-Ionized}). The second value is a result smoothing HI4PI to
a 0\fdg25 pixel grid similar to match the smoothed WHAM maps and was only applied to the ellipsoidal scenario.}

\tablenotemark{b}{Internal extinction correction for the SMC is only applied to \ha\ regions, which coincide spatially with \nhi\ $\geq 10.0^{20.0}$cm$^{-2}$. }

\tablenotemark{c}{No internal extinction values were available for the SMC filament
region.}

\end{table*}

To reduce the data, we used the ring-summing and flat-fielding procedures
described in Haffner et al. (2003). After the data were ring summed,
we calibrated the \ha\  velocities using atmospheric lines, subtracted
the atmospheric emission using an atmospheric template, and calibrated
the intensities to account for night-to-night differences in transmission
using $\lambda$ Ori as our standard intensity calibrator. These
procedures are further detailed below.

\subsection{Data Calibration}

After pre-processing the observations with the standard WHAM pipeline
\citep{haffner2003thewisconsin}, the spectra span a 200 \kms\ velocity
window with 2 \kms\ bins in an arbitrary velocity
frame. Figure \ref{fig:brightfaintcomparison} (i) highlights the
wing of the OH line, centered at a geocentric velocity of +272.44
\kms. This wing dominates the red edge of the
velocity window. We use the OH line in combination with an estimated
velocity shift based on multiple observations to find the velocity
offset. 

We assume that our background continuum level is linear over our 200
\kms\ window at all velocities. However, when
a foreground star is present, the spectra will be elevated and contain
stellar absorption lines. Beams that contain stars with $m_{V} < 6$
within a 0.55 radius ($\sim9\%$  of our observations) are excluded from this survey
to minimize this foreground contamination and are replaced with an
average of the uncontaminated observations within 1\arcdeg.

\begin{figure*}[t]
%\begin{raggedright}
\begin{tabular}{cc}
\includegraphics[bb=0bp 0bp 458bp 389bp,width=0.35\paperwidth]{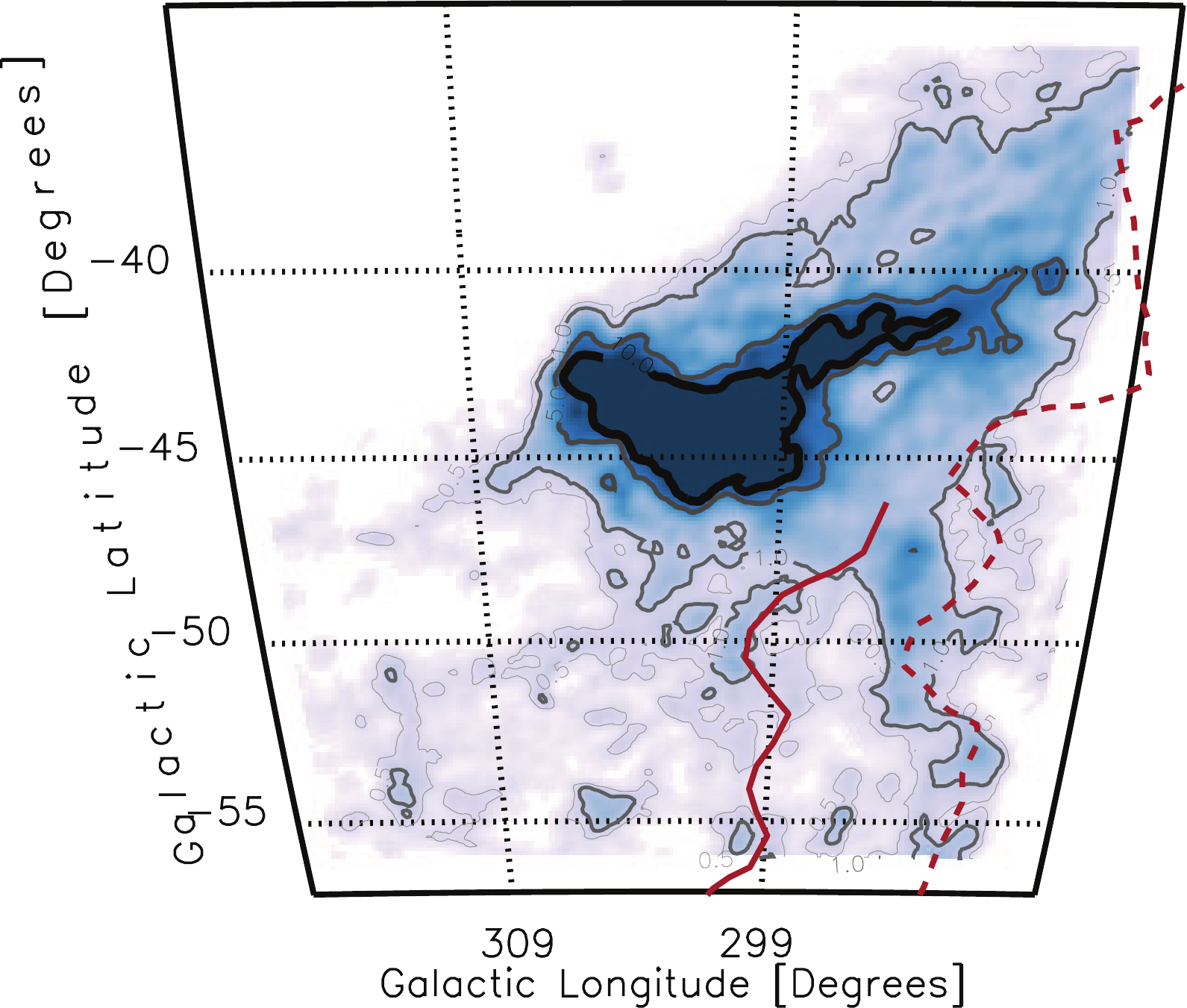} & \includegraphics[bb=0bp 0bp 458bp 389bp,width=0.35\paperwidth]{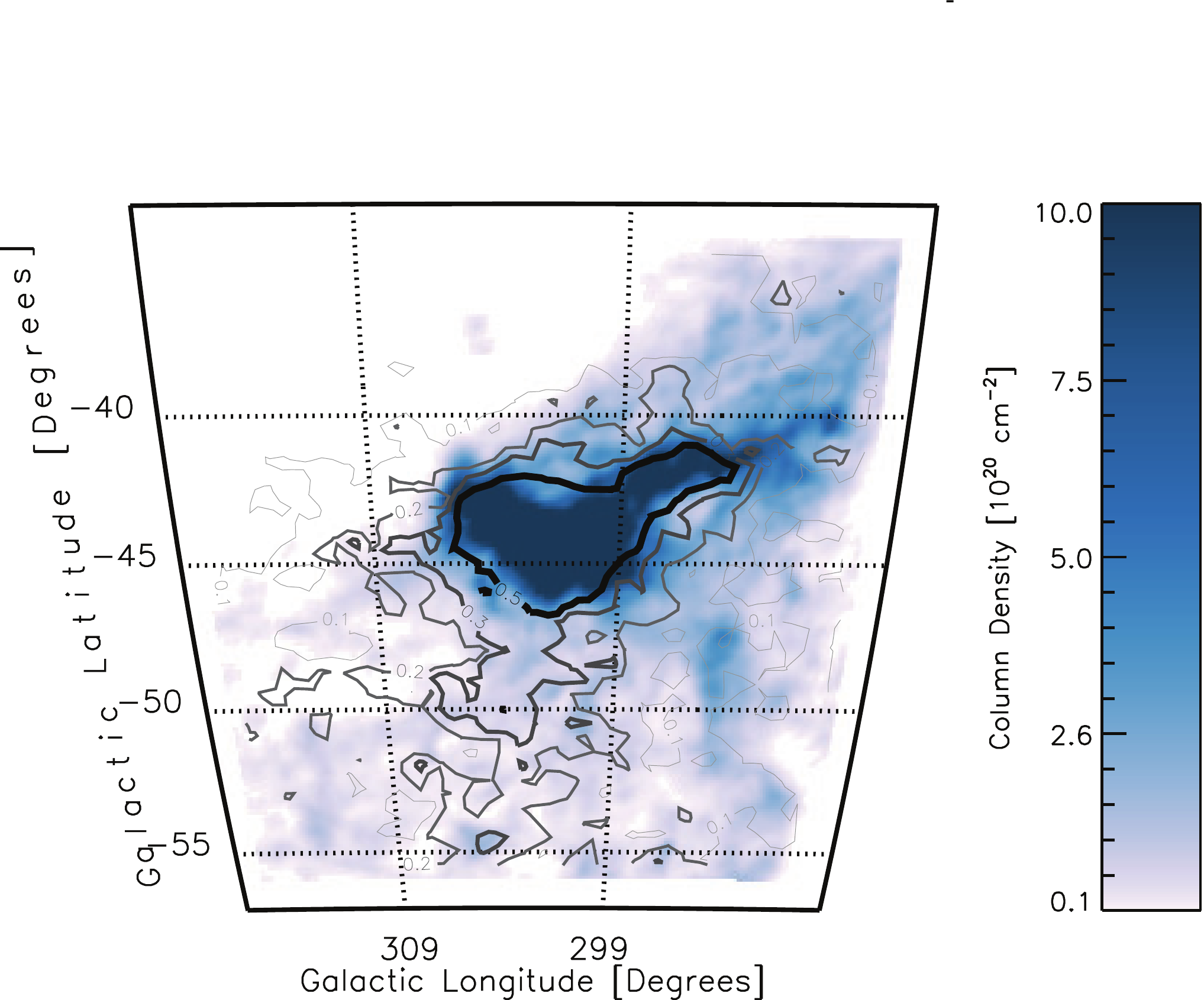}\tabularnewline
\includegraphics[bb=0bp 0bp 458bp 389bp,width=0.35\paperwidth]{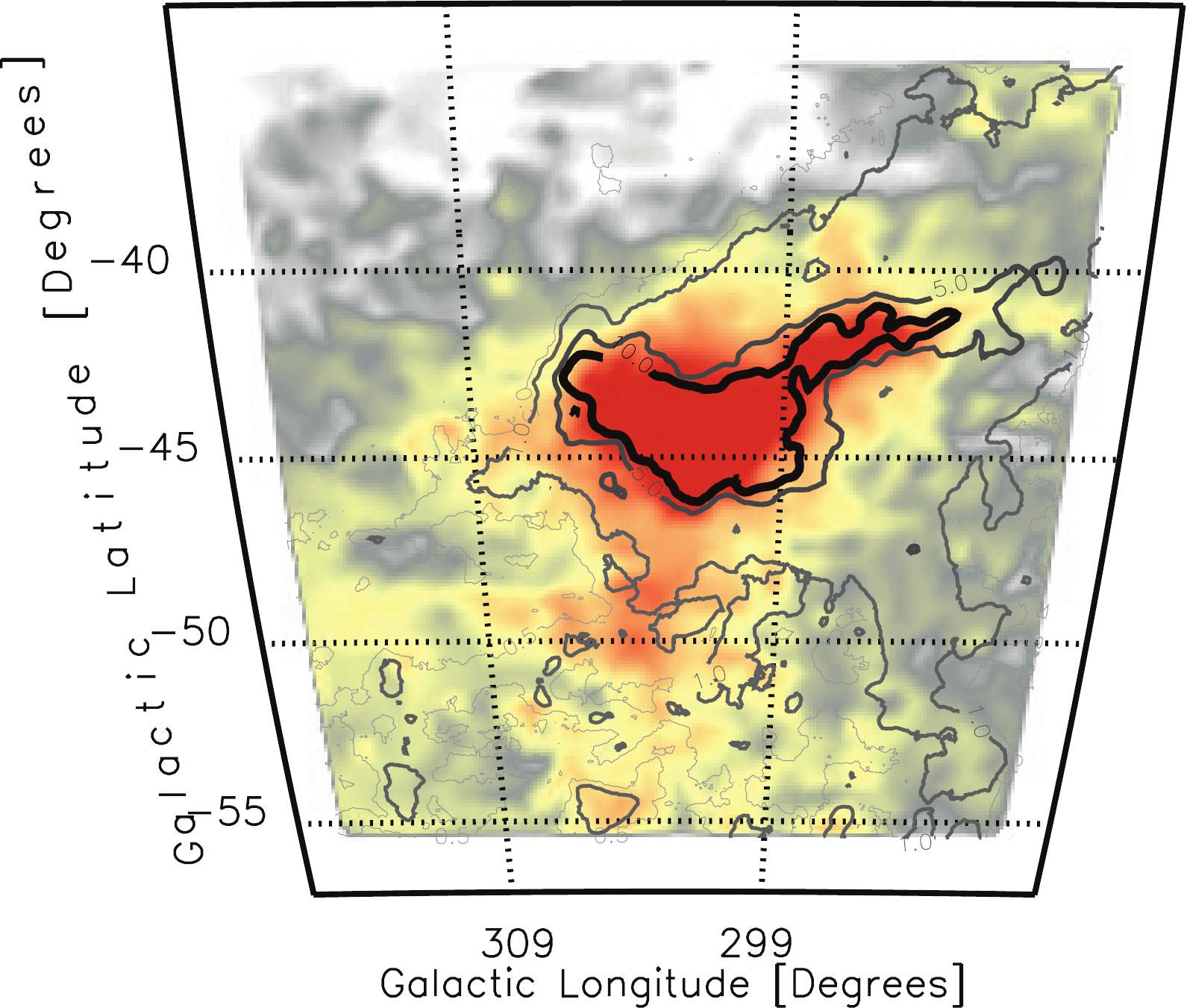} & \includegraphics[bb=0bp 0bp 458bp 389bp,width=0.35\paperwidth]{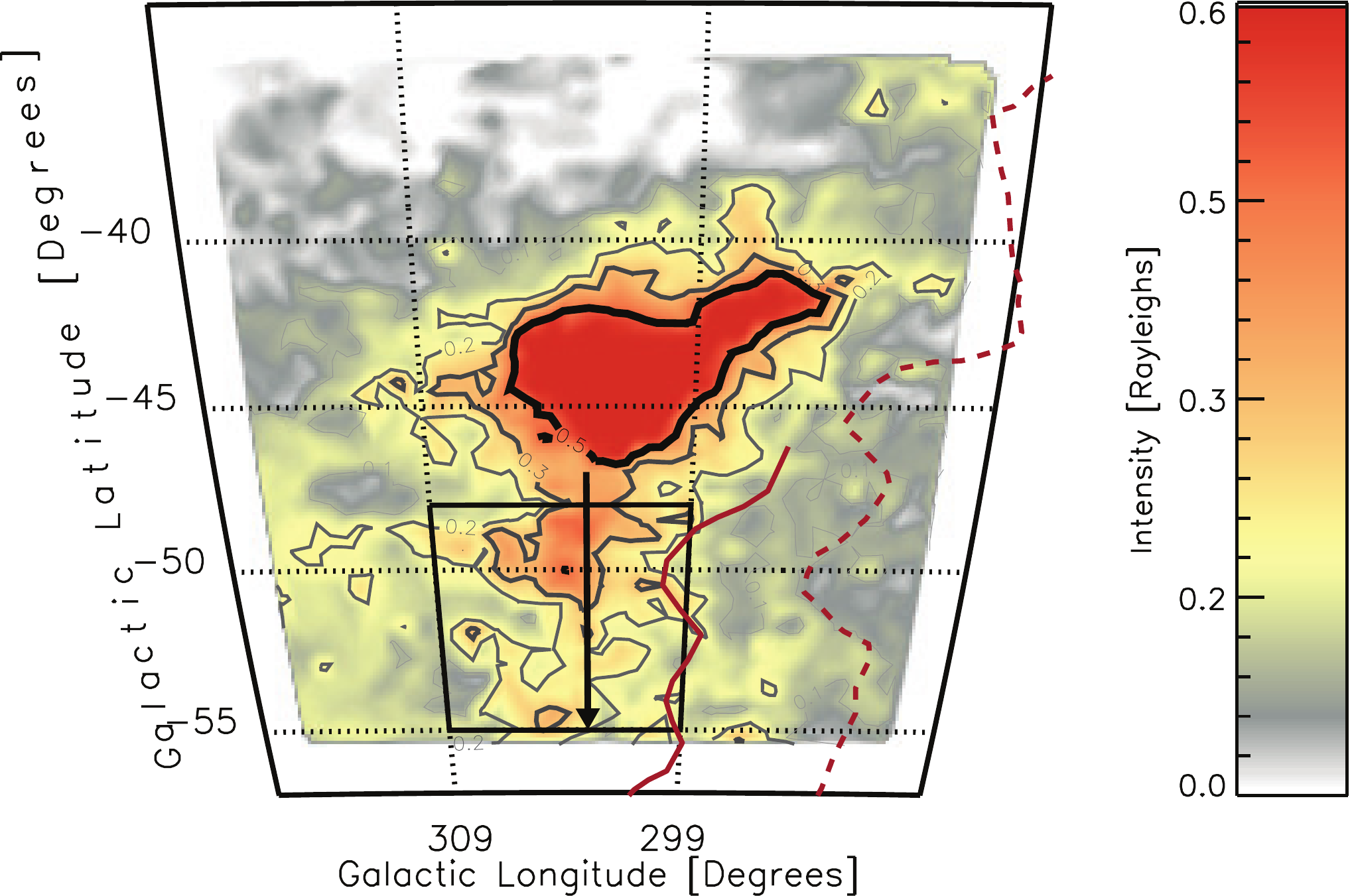}\tabularnewline
\end{tabular}
%\par\end{raggedright}

%\gridline{
%\fig{smc_hi_gass}{0.4\textwidth}{}
%\fig{smc_hi_gass_contour}{0.4\textwidth}{}}
%\gridline{
%\fig{smc_ha}{0.4\textwidth}{}
%\fig{smc_ha_contoured}{0.4\textwidth}{}}

\caption{Comparison of \hi\ (top row) and \ha\  (bottom row) emission
maps from the HI4PI survey and the WHAM survey respectively. 
The \ha\ intensity values are not extinction corrected. The
left column figures present the integrated emission maps with \hi\
contours overlaid, and the right column figures have the \ha\  contours
overlaid. The emission for both maps is integrated over $\vlsr = +90 \leq \vlsr \leq +210 \kms$.
The \hi\ scaling is clipped at a column density of $10^{21}$
cm$^{-2}$ to highlight the faint \hi\
emission. The \ha\  emission is clipped at 0.6 R. The solid red (SMC)
and dashed red (LMC) filaments mark the locations of the two \hi\
filaments within the Magellanic Stream identified by \citet{nidever2008theorigin}.
The black box marks the region that contains the SMC \ha\  filament,
and the back arrow marks the filament from $\lb = (303\fdg, -47\fdg0)$ to $(303\fdg0, -55\fdg0)$.
Due to scaling limits, the SMC filament does not appear clearly in
this projection. \label{fig:Comparison-of-HI}}
\end{figure*}

\subsection{Bright Atmospheric Line Subtraction \label{sub:Bright-Atmospheric-Line-Section}}

\begin{figure*}[tp]
%\begin{raggedright}
\begin{tabular}{cc}
\includegraphics[bb=0bp 0bp 459bp 418bp,width=0.35\paperwidth]{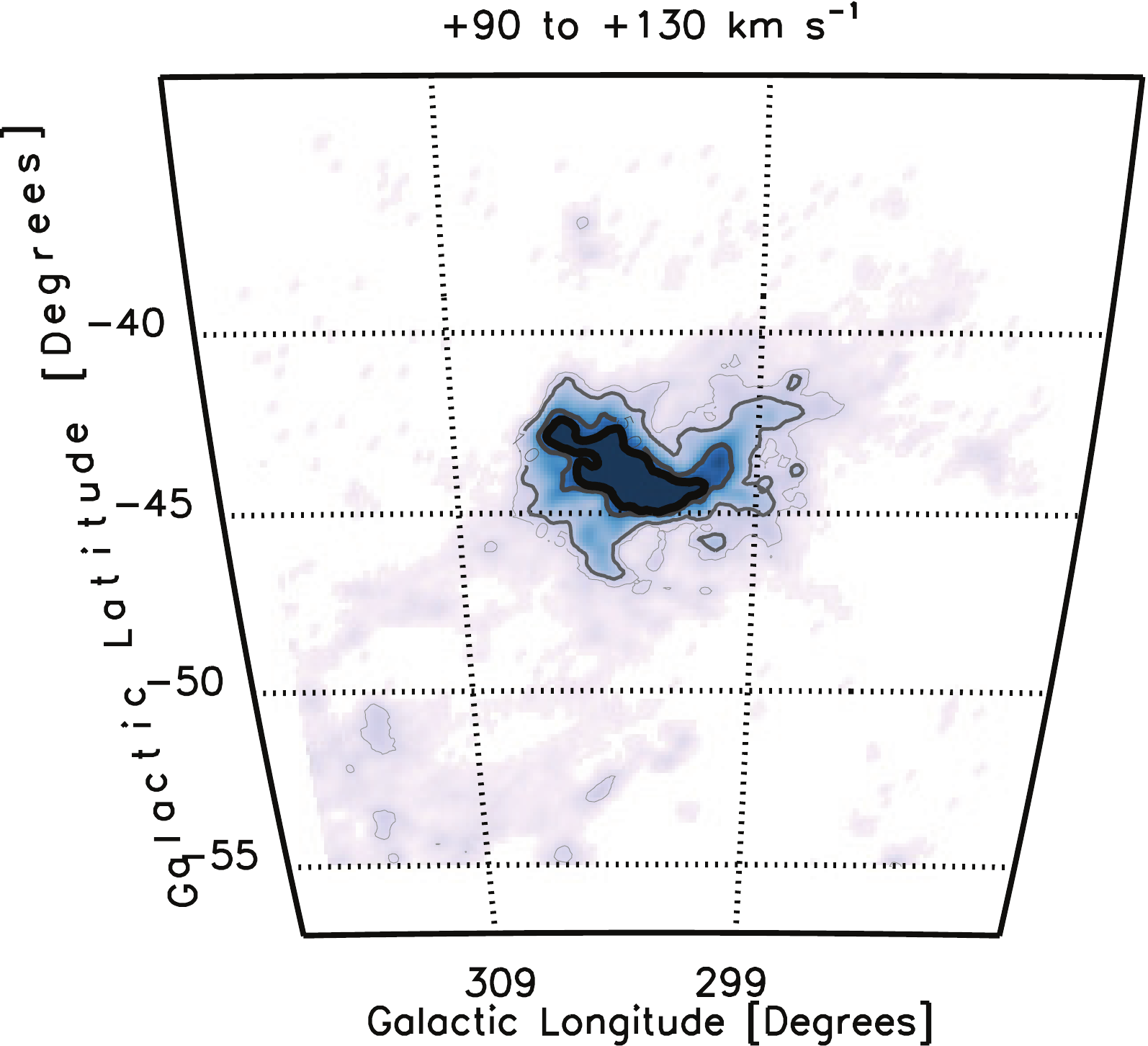} & \includegraphics[bb=0bp 0bp 459bp 418bp,width=0.35\paperwidth]{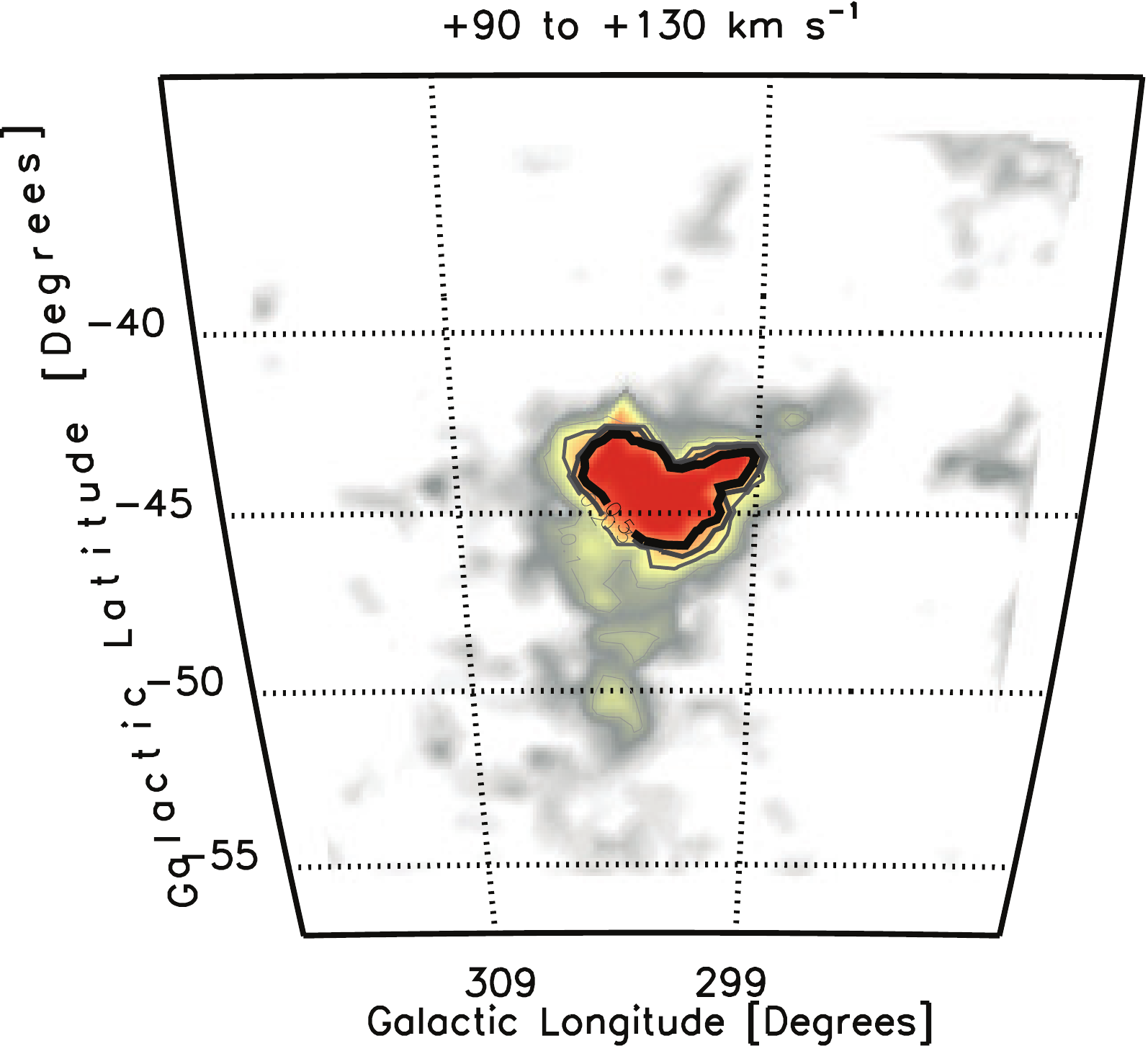}\tabularnewline
\includegraphics[bb=0bp 0bp 459bp 418bp,width=0.35\paperwidth]{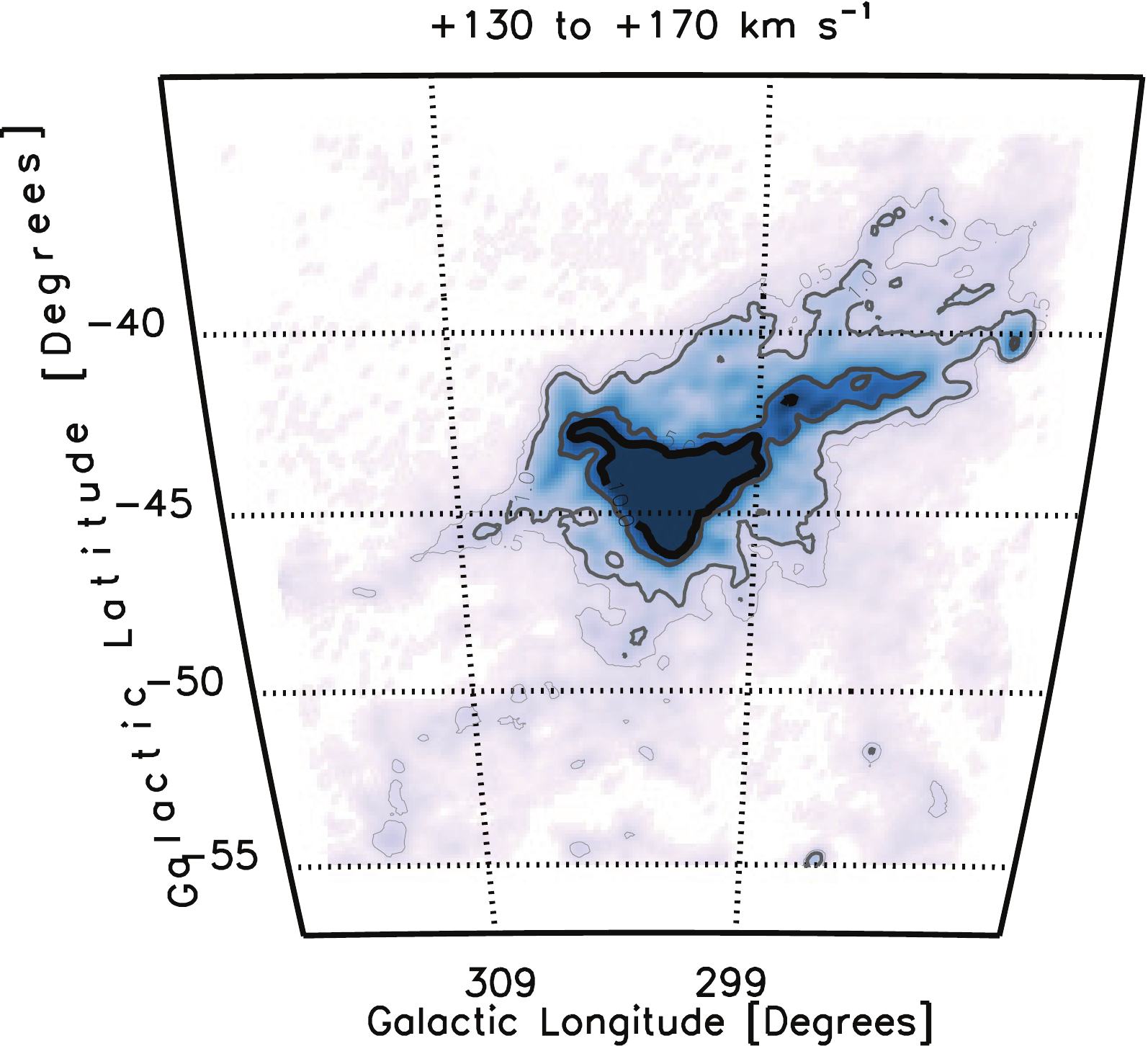} & \includegraphics[bb=0bp 0bp 459bp 418bp,width=0.35\paperwidth]{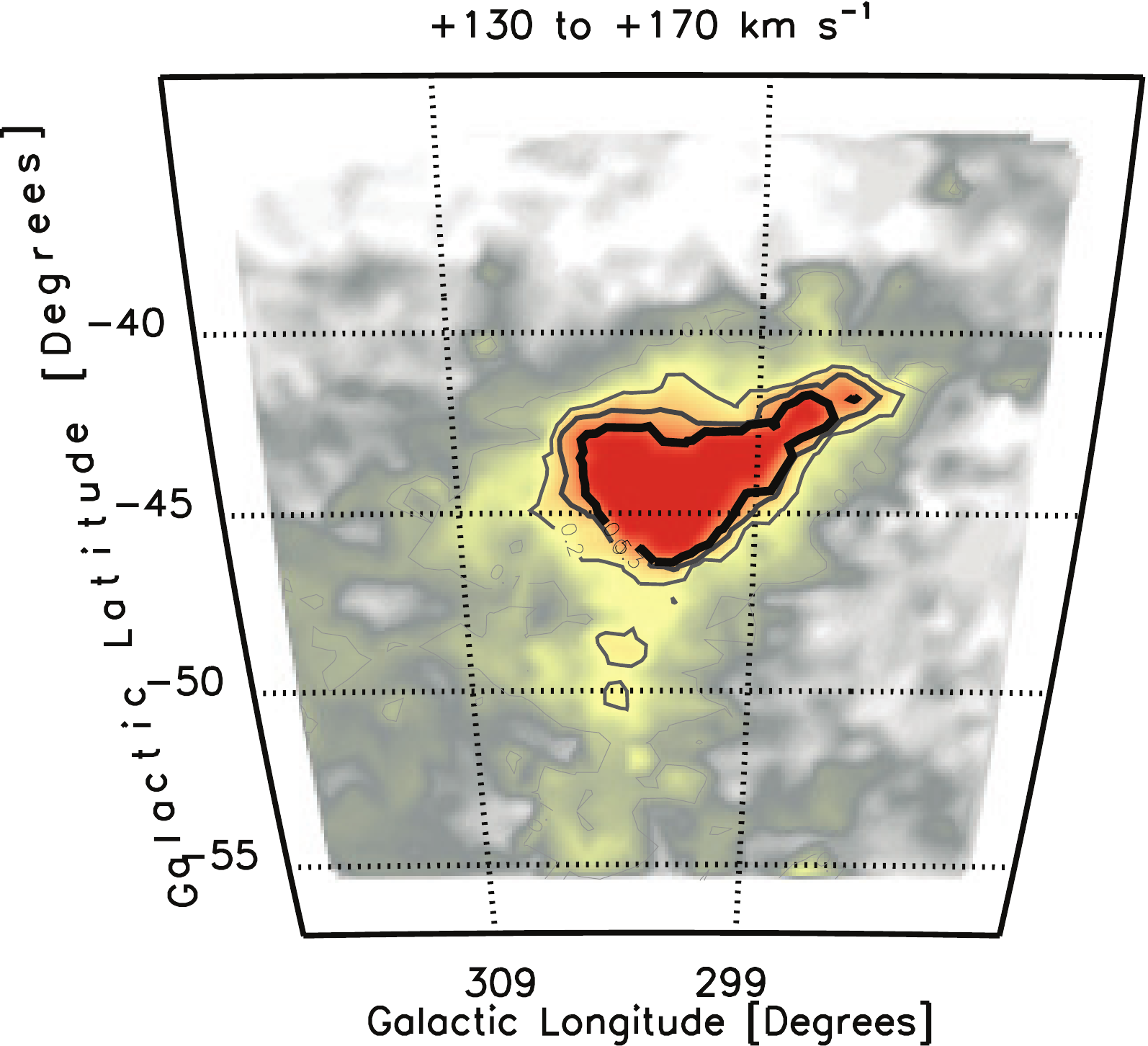}\tabularnewline
\includegraphics[bb=0bp 0bp 484bp 496bp,width=0.35\paperwidth]{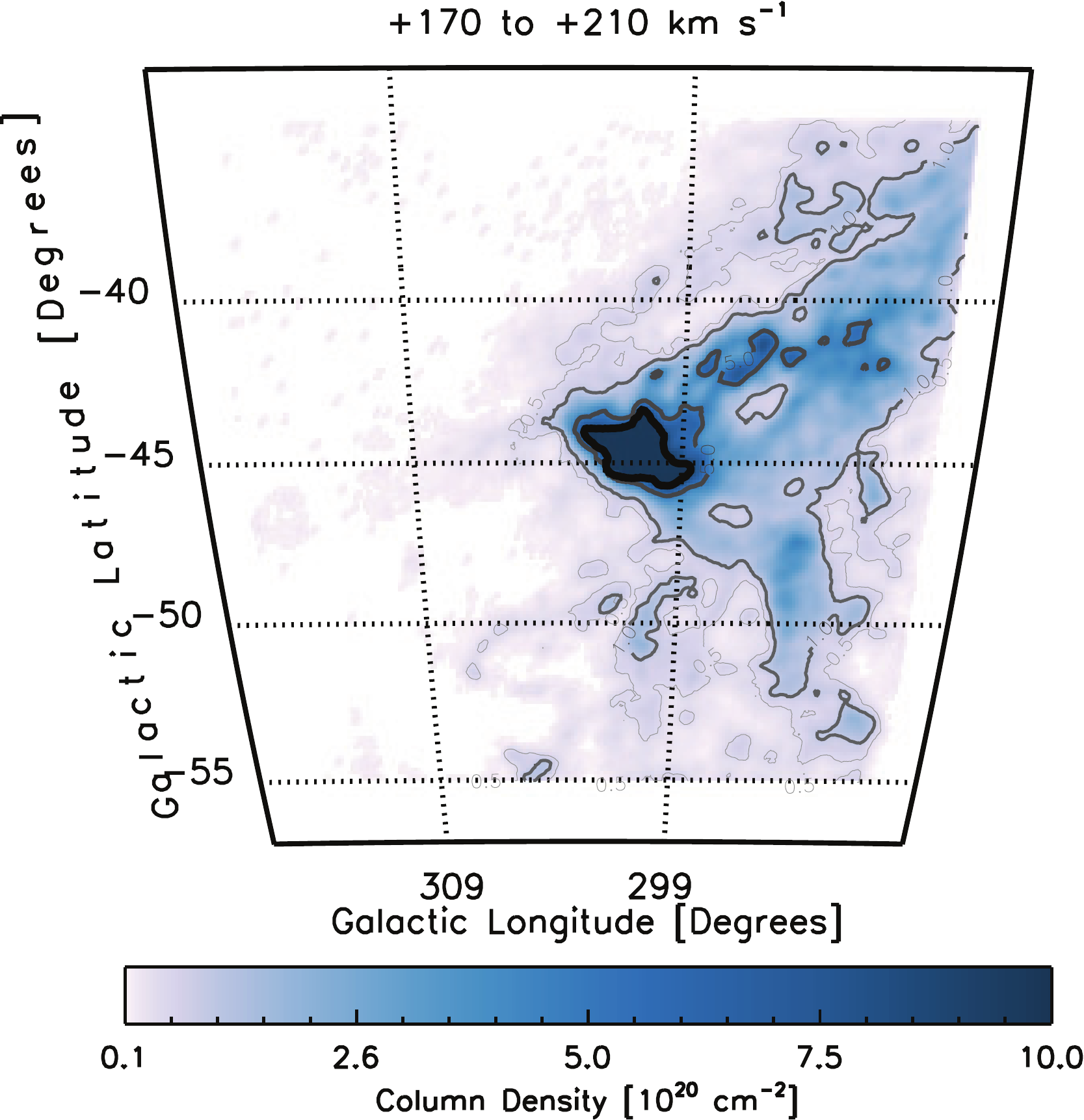} & \includegraphics[bb=0bp 0bp 480bp 496bp,width=0.35\paperwidth]{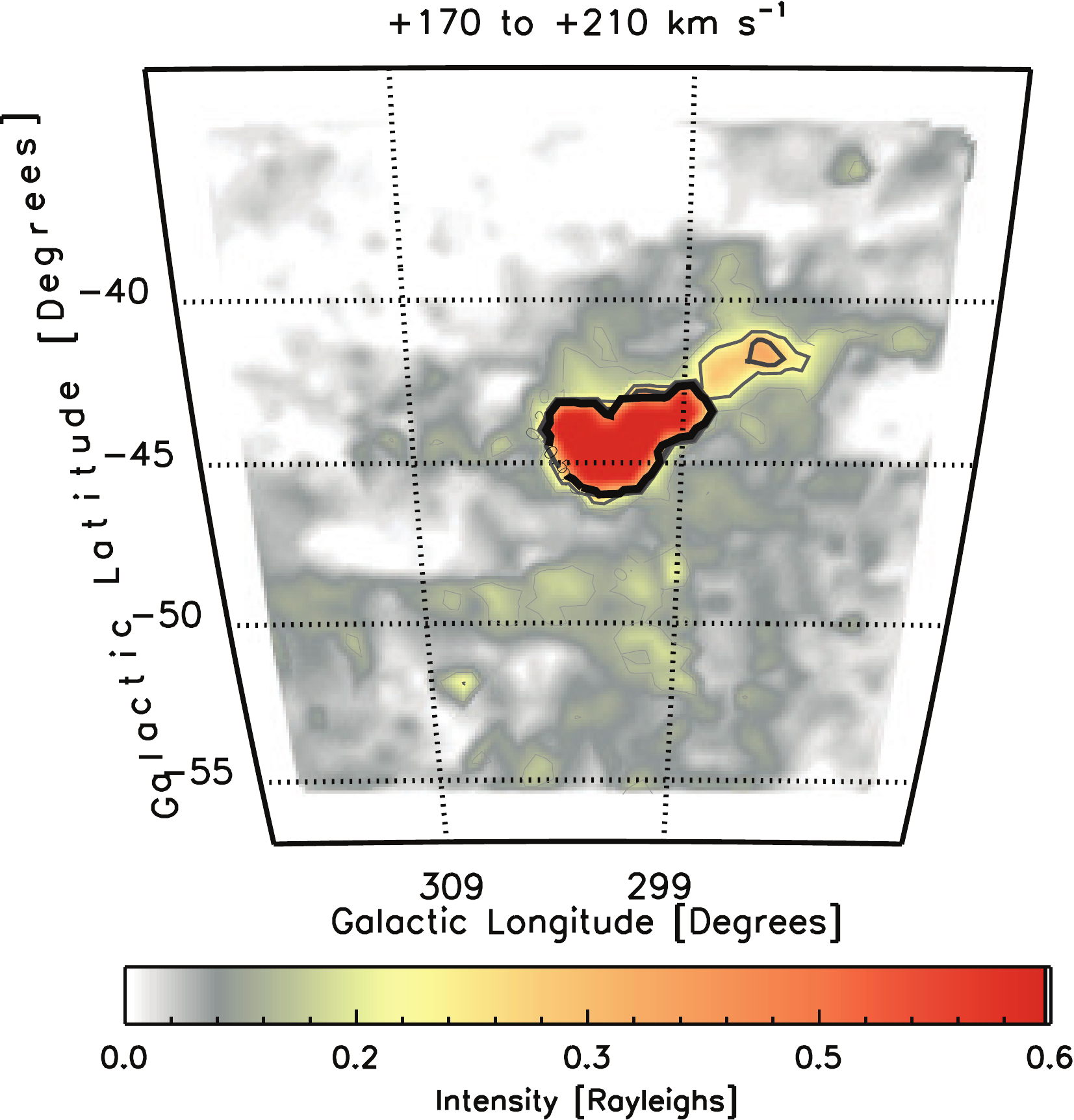}\tabularnewline
\end{tabular}
%\par\end{raggedright}

\caption{Comparison of \hi\ (left column) and \ha\  (right column)
emission maps from the HI4PI survey and the WHAM survey at different
velocity channels. The \hi\ is more compact in each channel
map, while the \ha\  is far more extended across all velocities.
Much of of the diffuse \ha\  extends far beyond the observable boundaries
of the SMC's \hi\, with less structural coherency than the
neutral gas. \label{fig:VelSlices}}
\end{figure*}

Night-to-night changes in the atmosphere cause variations in the strength
of the bright and faint atmospheric lines. The OH line \citep{1989JGR....9414629M}
is produced from interactions between solar radiation and Earth's
upper atmosphere and varies in strength with the direction and the
time of the observation based on the Earth-Sun geometry. Therefore,
we always fit the OH line at $v{}_{\text{geo}}$ = +272.44 \kms\
separately from the faint atmospheric lines, which vary due to the air mass
(see Section \ref{sub:Faint-Atmospheric-Line-Section}). There are
two main sources that alter the shape of the bright atmospheric lines: (1)
The ``tuning'' of the dual etalons, where the precision in aligning
the dual-etalon transmission functions can have slight night-to-night
variations in the instrument profile. These effects are only detectable
in narrow, bright lines. (2) A geocoronal ``ghost''.
The ``ghost'' lies under the lies underneath the OH line at $v_\textrm{geo} = +272.44$ \kms. As described
in \citet{haffner2003thewisconsin}, it is the result of incomplete
suppression of the geocoronal line at $v_\mathrm{geo} = -2.3$ \kms\ from a neighboring order in the
high-resolution etalon. In the velocity window observed in this study
($+50 \leq \vlsr\ \leq +250$ \kms),
some of the SMC emission may lie under the OH wing. However, because
the velocity in this window only contains a portion of the OH line,
it is difficult to  fit well. To minimize over-subtracting the SMC
emission, we assign a width of 10 \kms\
to the narrow OH line  Gaussian fit to prevent inclusion of lower velocity emission not
attributed to the OH line. However, removal of some higher velocity
SMC emission is unavoidable. We therefore integrate our emission maps
over a $+90\ \leq \vlsr \leq +210\ \kms$ to avoid the contaminated regions. All \hi\  
emission maps and mass calculations are integrated over the same range for consistency. %

%Our observations also include fainter atmospheric emission below \ensuremath{\sim}
%0.1 R at all velocities (3.4). The removal of both the bright and
%faint atmospheric lines is crucial for detecting the faint \ha\
%emission between the Magellanic Clouds. 

\subsection{Faint Atmospheric Line Subtraction\label{sub:Faint-Atmospheric-Line-Section}}

While the OH line dominates over the other atmospheric emission, faint
atmospheric lines are present throughout the entire spectrum with
intensities of $~{\sim0.1}$~{R}. The strength of these lines is correlated
with the air mass. \citetalias{barger2013warmionized} created a synthetic atmospheric spectra to describe
the faint atmospheric lines observed at CTIO by observing two faint
directions multiple times over a 10 day period. Properties of the
atmospheric lines can be found in table 1 and figure 3 of \citetalias{barger2013warmionized}, and
the template creation and subtraction process is also described by
\citet{hausen2002interstellar} and \citet{haffner2003thewisconsin}.
Our averaged spectrum consists of numerous 30 and 60 s observations,
totaling 4.5 hours, toward $(\ell, b) = (60\fdg0\arcdeg,  -67.0\arcdeg )$
and $(89.0\arcdeg, -71.0\arcdeg)$. These are ``faint''
directions that appear to have no strong \ha\ emission,
so the atmospheric lines can be identified. To account for changes
in the flux due to the air mass and daily fluctuations, we scale the synthetic
template to match the emission present in a high signal-to-noise ratio block-averaged
spectrum. We then subtract the scaled atmospheric template from the
observation. The total integrated intensity of the template over the
SMC window is $\sim$0.07 R.

\subsection{Extinction Correction}

\subsubsection{Foreground Extinction}

The \ha\  intensity emitted from the SMC is attenuated by foreground
dust from the MW, as well as dust from within the SMC itself. To correct
for foreground MW extinction, we used the method outlined in \citetalias{barger2013warmionized} ) their section
3.4.1):

\begin{equation}
I_{H\alpha,corr}=I_{H\alpha,obs}e^{A(H\alpha)/2.5}
\end{equation}
where,

\begin{equation}
A(H\alpha)=5.14\times10^{-22} \left\langle \nhi \right\rangle\ \textrm{cm} ^{2}\ \textrm{atoms}^{-1} \ \textrm{mag}
\end{equation}

We integrate the total \hi\ intensity over the $-100$ to $+100$
\kms\ velocity range. The extinction correction is important to 
the mass calculations in Section \ref{sub:Mass-of-Ionized}. We use three different methods to estimate the mass.
In one, called the ellipsoidal scenario, we calculate the extinction per beam. In the other two, we use the extinction averaged over the area
to stay consistent with the mass calculation methods. Accounting for the dust attention due to the foreground materials results in a 
12.9\%--14.4\%
increase in \iha\ resulting from the foreground
extinction for the SMC region, a 14.6\%--16.6\% and 13.7\%--15.8\%  increase in the
\hi\ and \ha\  Tail regions,
and a 6.4\%--7.6\% increase from the SMC Filament foreground (Table
\ref{tab:Extinction}).

\subsubsection{SMC Extinction}

The extended envelope of the SMC has not been thoroughly studied for
dust content. However, the bar of the SMC was sampled in \citet{Gordon2003}.
They find an $R_{V} = 2.74, A_{V} \backsimeq 7.6 \times 10^{-23}\ \langle \nhi \rangle$ cm$^{2}$ mag
and an average $E (B-V) = 0.179$ mag. To estimate the \ha\  extinction
in the SMC, we combine equations 1 and 4 from \citet{Gordon2003}
to describe how the dust-attenuated \ha\  emission scales with \nhi:

%\begin{equation}
%A(H\alpha)=1.057\times10^{-22}\left\langle N_{\hi\}\right\rangle \text{cm $^{-2}$\ensuremath{\cdot}atom\ensuremath{s^{-1}\cdot}mag}\label{eq:ExtincCenter-1-1}
%\end{equation}

\begin{equation}
A(\textrm{\ha})=\left({\frac{E(\textrm{\ha} - V)}{{E(B - V)}}}\right){R_V}^{-1}\langle N_\textrm{\hi}\rangle\ \textrm{cm}^{2}\ \textrm{mag}  \label{eq:ExtincCenter-1}
\end{equation}

We calculate the value using results listed in \citet{Gordon2003} Tables 2 and 3 for the
SMC, $E(\ha - V)$ = 0.197 mag, and the resulting
equation is $A(\ha)=1.057\times10^{-22}\left\langle \nhi \right\rangle\ \textrm{cm}^{2}$ mag.
We apply the extinction correction only on the central region of the
SMC for observations (or beams) with an \nhi\ 
$>1.0\times10^{20}$ cm$^{-2}$  over an integrated
 velocity range of $+90 \leq \vlsr\ \leq +210$~{\kms}, where the gas is coincident
with the visible stellar component of the galaxy. \nhi\
is determined using the HI4PI survey smoothed to
0.25\arcdeg{} pixels to match our angular resolution when calculating the
extinction in our ellipsoidal scenario \citep{HI4PI2016}. Our ionized
skin and cylindrical scenarios 
use the average \nhi\ of the region to
calculate the extinction for the regions in the SMC where  \nhi\  $>1.0\times10^{20}$ cm$^{-2}$.
All areas of the SMC with lower column than our cutoff were only foreground-extinction corrected.
Correcting the \ha\ intensities
for the dust within the structure results in an average 21--23\%
increase for the  region of the SMC where $\iha > 0.5$~R.
 Because the \ha\
emitting regions lie throughout the SMC and not necessarily behind
the full dust extent, our correction provides an upper limit for the
\ha\ intensities. 

%\begin{equation}
%A(H\alpha)=2.05\times10^{-22}\left\langle N_{\hi\}\right\rangle \text{cm $^{-2}$\ensuremath{\cdot}atom\ensuremath{s^{-1}\cdot}mag}\label{eq:Tail-1}
%\end{equation}

\subsection{\ha\ Tail and \hi\ Tail}

\citetalias{barger2013warmionized} included
 an analysis of the \ha\ and \hi\ Tail of the SMC (see the region designated in Figure \ref{fig:Skematic}). 
 Based on their WHAM observations, the region was integrated over $+100 \leq \vlsr \leq\ +300$ \kms. 
 They calculated a mass estimate using the same methods described in Section \ref{sub:Mass-of-Ionized}. 
We use an updated data reduction and intensity
correction method compared to the 2013 paper. The previous WHAM reduction
pipeline included a standard 0.93 intensity correction applied universally
to each night. The new pipelines calculates a night-to-night intensity
variation based on the seeing conditions for a night and uses a standard
calibrator ($\lambda$ Ori) to calculate the transmission. In addition
to the night-to-night variation, we include the transmission degradation
 WHAM optics in our corrections. On average, this results in
a $\sim20\%$ correction to the intensity used in the \citetalias{barger2013warmionized}.

\section{\ha\ Intensity map \label{section:Intensity}}

We observed the SMC over $+50\leq \vlsr \leq +250$ \kms.
Figure \ref{fig:Comparison-of-HI} displays the non-extinction-corrected
total integrated intensity of the SMC in \ha\  and a column density
map of \hi. The \hi\ and \ha\  map are integrated
over $+90 \leq \vlsr \leq +210$ \kms\ to avoid the wings
of the OH line near $v_{\rm geo} = +272.44$ \kms
and MW emission at $\vlsr <+90$ \kms.
The sensitivity of our survey is $\iha \backsimeq 10$
mR, for extended continuous emission, and $\iha \backsimeq30$
mR for individual observations. The \hi\ spectra were obtained
from the HI4PI survey and have a sensitivity
of $N_{{\rm H}\textsc{~i}}=2.3\times10^{18}\text{cm}^{-2}$ for line of 1.49 \kms\ wide \citep{Bekthi2016}. 

For the center of the galaxy, there is a strong correlation between
\ha\  emission and $N{}_{{\rm H}\textsc{~i}}$. The brightest \ha\ 
emission in the SMC traces closely the regions where \nhi\
is greater than $10^{20} \textrm{cm}^{-2}$,
with particularly strong correlation between bright \ha\ emission
and \nhi\ of $5\times10^{20} \textrm{cm}^{-2}$
and above. When compared to MCELS (Figure \ref{fig:MCELS}), we see
the strongest emission correlates with the stellar component of the
galaxy that MCELS had previously observed. 

\subsection{Distribution}

%\subsection{\hi\ and HA average width}
%
%Something is wrong with the average width calculations. Needs to be
%fixed. These global trends are shown in Figure \ref{fig:MeanVelocity}. 

Outside the core of the SMC, the brighter \ha\  emission does not
appear to directly correlate with the densest \hi\  regions.
Toward the Stream, there is a bright region of \ha\  emission starting
at $(\ell,b)= (305.5\arcdeg , -50.0\arcdeg )$ in a region
we will refer to as the SMC-Stream interface. Conversely, the dense
neutral filament from $b = -47\arcdeg{}$ to $-55\arcdeg{}$
has little \ha\ emission. The faint \ha\  emission reaches
several degrees beyond the boundaries of the \hi\ gas to
the south, extending out toward the Magellanic Stream. \ha\  emission
above 0.16 R can be observed in areas without bright \hi\
emission, such as the \hi\ gap from $(\ell,b) = (315\fdg5 ,
-50\fdg0 )$ to $(307\fdg0 , -47\fdg5)$.  While WHAM is reliably sensitive
 to emission lines at \iha $\geq 25$ mR, WHAM can detect continuous emission above 10 mR.

In Figure \ref{fig:VelSlices}, we present $N{}_{{\rm H}\textsc{~i}}$
and $I_{{\rm H}\alpha}$ maps at several velocity ranges, $+90
\leq \vlsr \leq +130$ \kms, $+130 \leq \vlsr \leq +170$ \kms,
and $+170 \leq \vlsr \leq +210$ \kms. The channel maps show
the wide distribution of \ha\  emission at several velocities in
the same spatial location. In contrast, the \hi\ channel
maps show the neutral component stays coherent in both velocity space
and location. 

\begin{figure*}[tb]
%\begin{centering}
%\includegraphics[bb=0bp 50bp 1448bp 480bp,width=0.8\paperwidth]{smc_spur_figure}\tabularnewline
\includegraphics[width=\textwidth,trim=0 70 0 0]{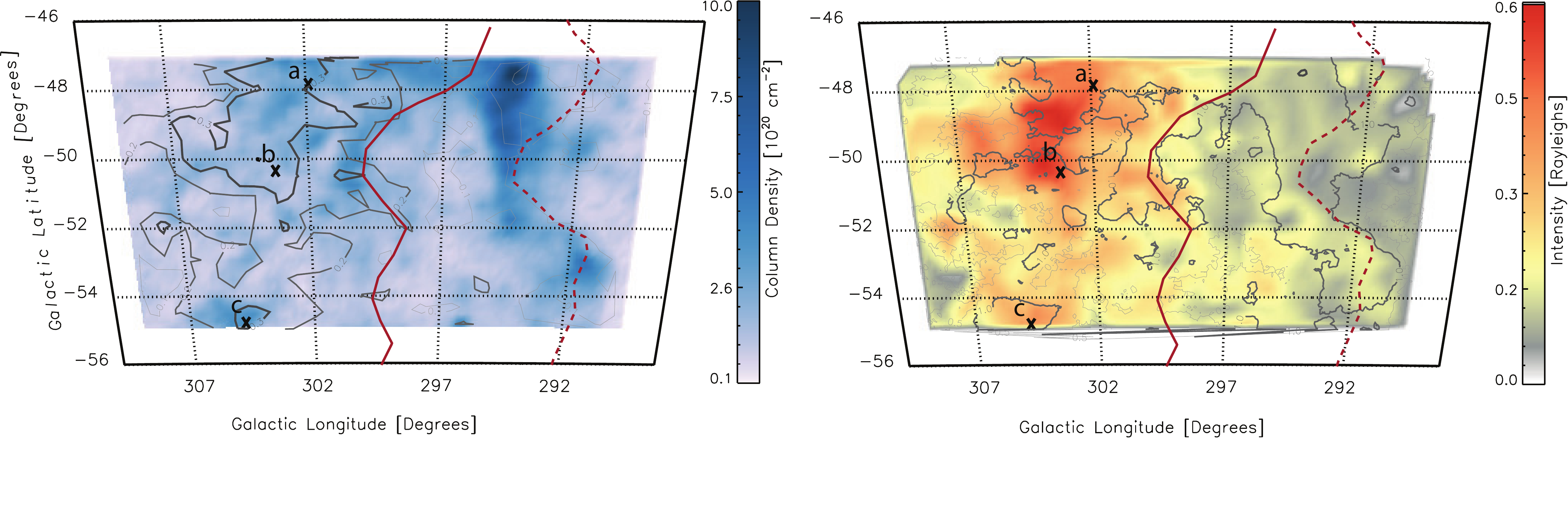}
%\par\end{centering}

\caption{SMC \ha\  Filament: \hi\ (left) emission with \ha\  contours,
and \ha\  (right) emission with \hi\ contours overlaid.
The filament runs along l = 305\arcdeg{} and extends from b =-47\arcdeg .
The solid red line traces the SMC \hi\ filament and the
dashed red line marks the LMC \hi\ filament. Both of
these filaments feed gas into the Stream and were originally kinematically
traced by \citet{nidever2008theorigin}. The \ha\  filament appears
to mirror the \hi\ SMC filament, but does not have a strong
\hi\ counterpart. The
x's marked a, b, and c indicate the location of the emission fit in
Figure \ref{fig:FilamentSpectra}. \label{fig:HIFilamentcomparison-1}}
\end{figure*}

\begin{figure}[tb]
%\begin{centering}
%\begin{tabular}{c}
\includegraphics[clip,width=1\columnwidth]{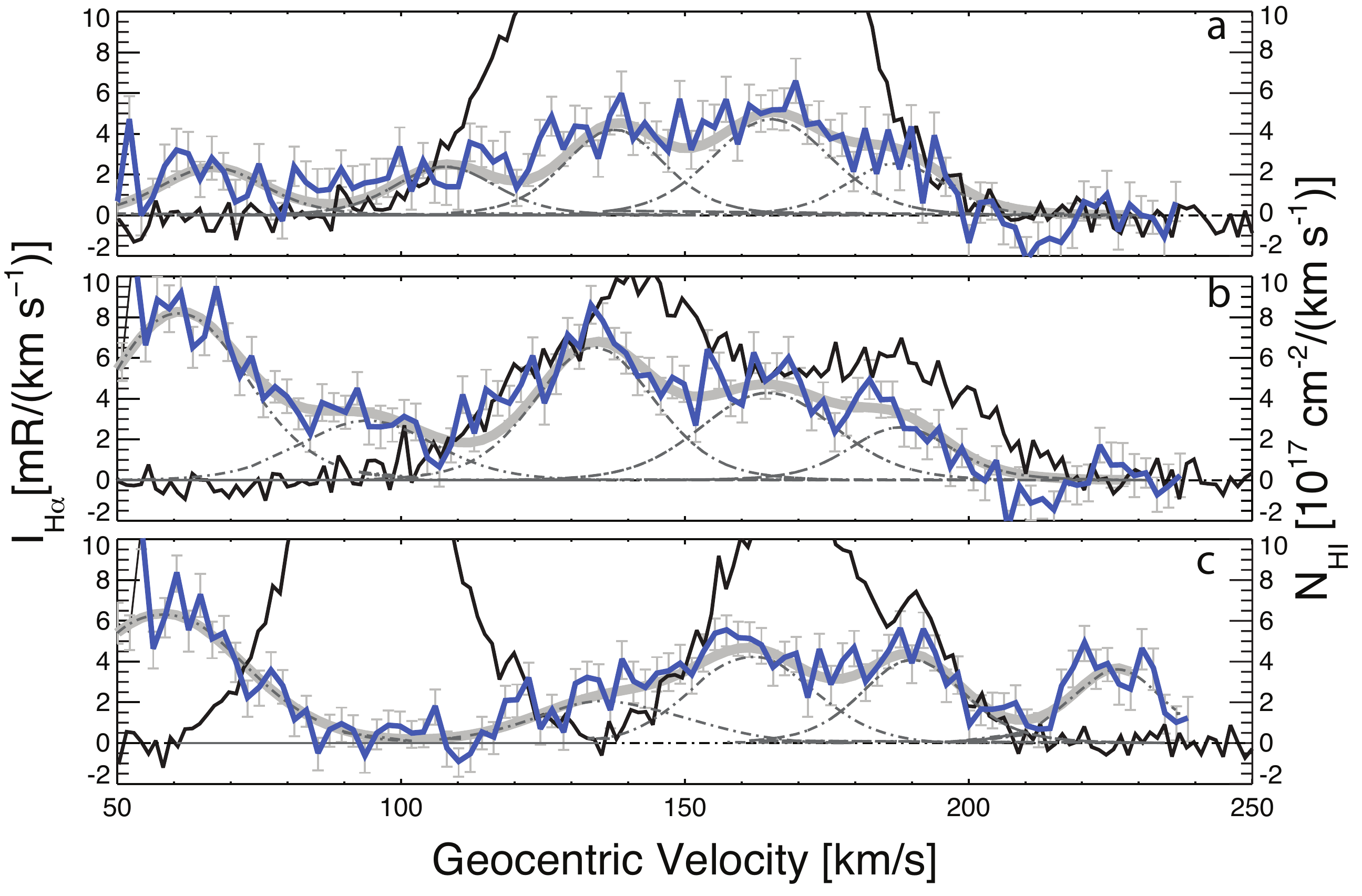}\tabularnewline
%\end{tabular}
%\par\end{centering}

\caption{Three pointings from within the SMC filament. The blue lines show
the \ha\  emission observed by WHAM, and the black lines marks the
\hi\ emission from the HI4PI survey. The dark gray contour
marks the Gaussian fit to the \ha\  emission and the dashed Gaussians
outlines the induvidually fit Gaussians. The location of each Gaussian
is marked in Figure \ref{fig:HIFilamentcomparison-1}. \label{fig:FilamentSpectra} }
\end{figure}

We find a  region of ionized gas extending out from the SMC,
which is marked by the black box in Figure \ref{fig:Comparison-of-HI}
(bottom right). This region begins at $(\ell,b) = (303\fdg0 ,
-47\fdg0)$ and extends down to the edge of the observed
region at (303\fdg0 , -55\fdg0 ). The ionized filament
is marked by the black arrow. This ionized filament runs parallel
to two neutral filaments discovered in \citet{nidever2008theorigin},
marked by the solid red and dashed red lines. Figure \ref{fig:HIFilamentcomparison-1}
focuses on the region of the extended SMC containing the ionized SMC
filament, with several spectra from the region in Figure \ref{fig:FilamentSpectra}.
While there is detectable \ha\  on the order of 0.1 R along the \hi\
filament marked by the solid red line, similar emission is also present
in the extended region beyond the filament and does not appear directly
related to the \hi filament. In contrast, along $\ell=305\arcdeg$
ionized filament, there is a \ha\  enhancement, with intensities
of 0.2 -- 0.5 R. This filament is spatially coincident with several
\hi\ clouds with column densities of $(0.5-1.0)\times10{}^{20}\textrm{cm}{}^{-2}$.
However, no extended filamentary structure comparable to the \hi\
filament exists at lower longitudes. 

%The SMC filament does not appear as strong as the LMC counterpart,
%however it is mirrored by nearby \ha\  emission. 

\subsection{\hi\ and \ha\  Velocity Distribution}

\begin{figure}[tb]
\begin{tabular}{r}
\includegraphics[bb=0bp 0bp 586bp 390bp,width=0.9\columnwidth]{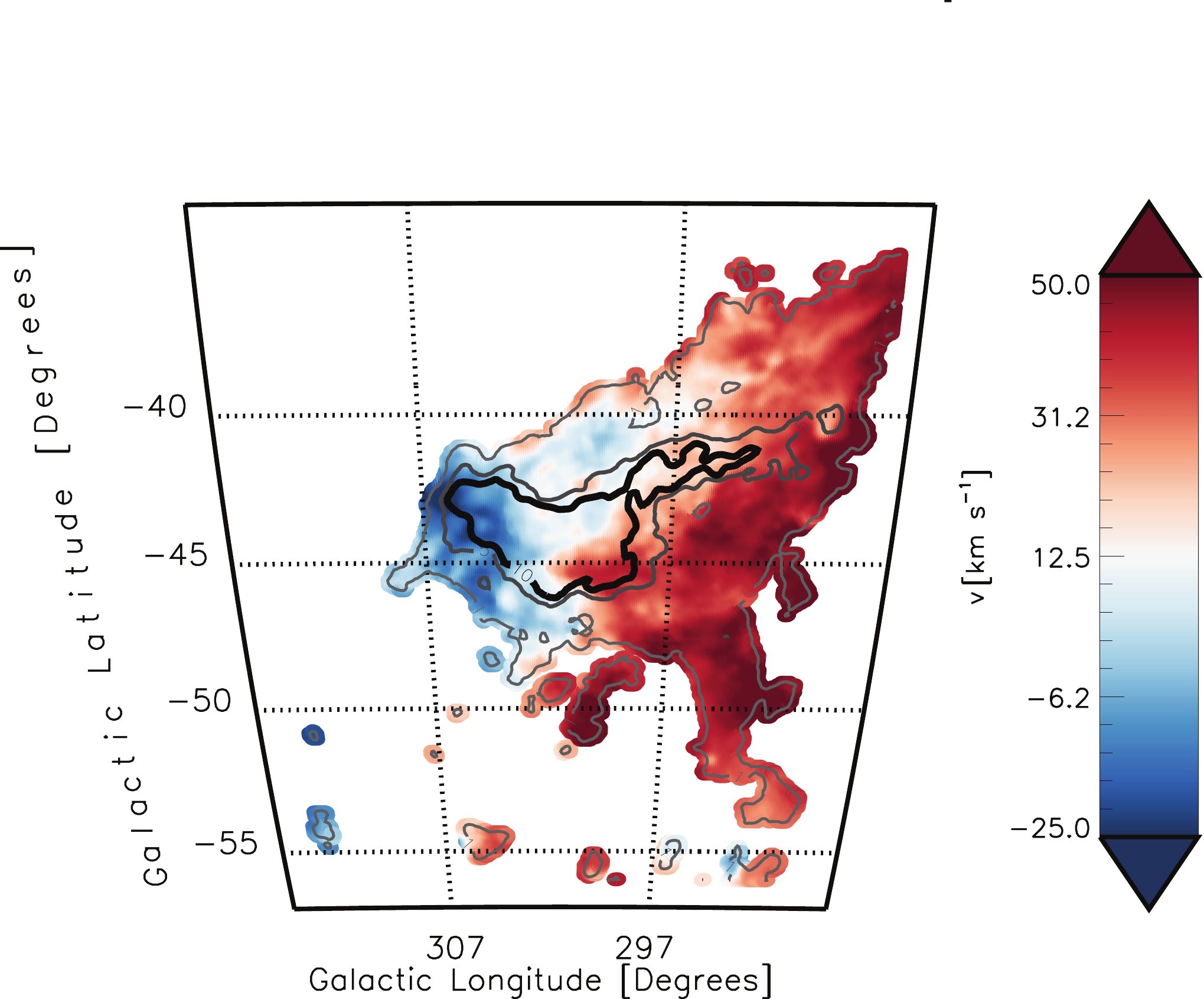}\tabularnewline
\includegraphics[bb=0bp 0bp 586bp 389bp,width=0.9\columnwidth]{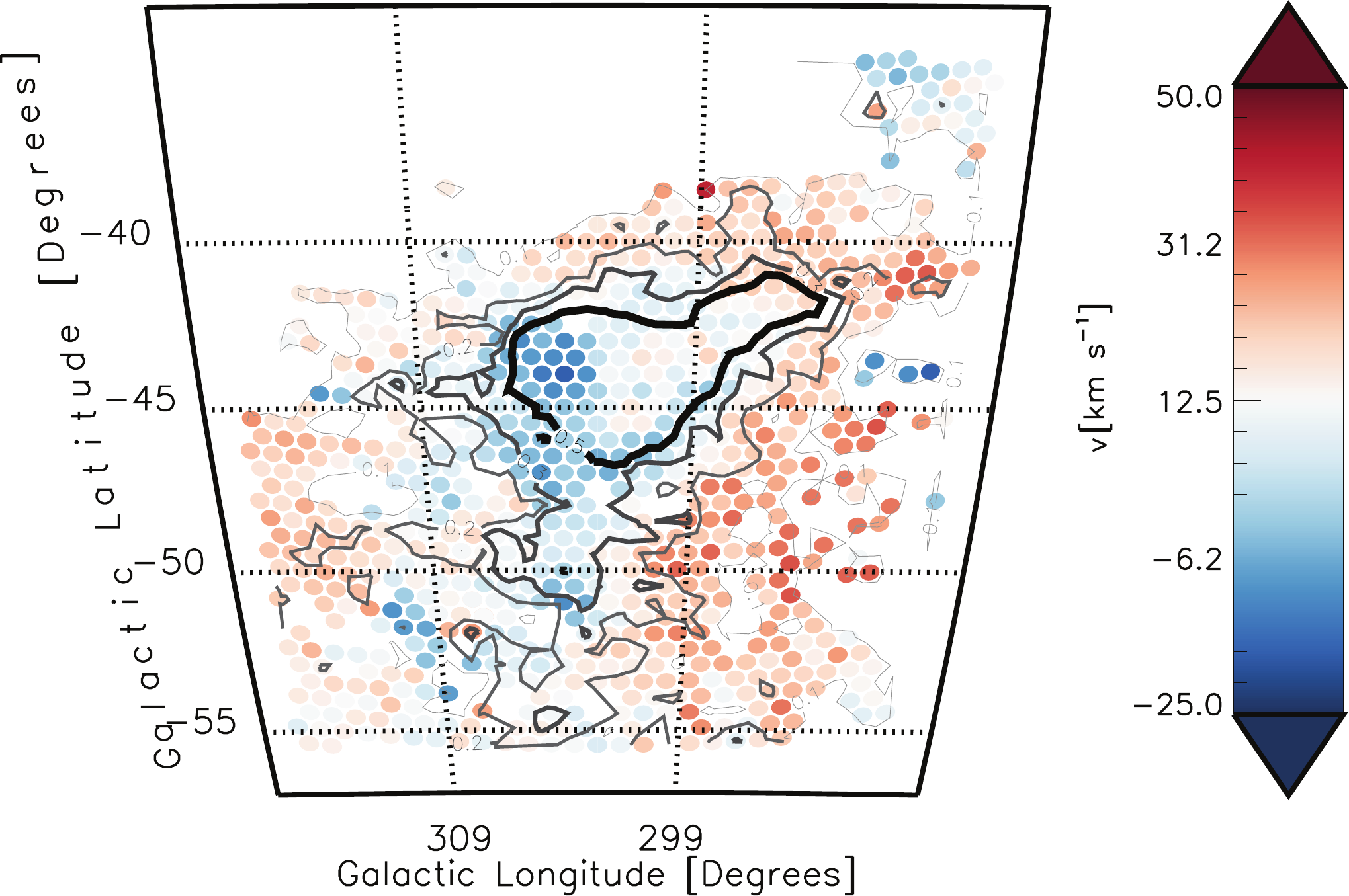}\tabularnewline
\end{tabular}

\caption{First Moment Map: The mean velocities of the gas traced by the \hi\
(\emph{top}) emission and the \ha\ (\emph{bottom}) emission. Velocities are centered around $\vlsr = +145 \,\kms$. Rotation of the  ionized gas,
while weaker, appears to follow the same axis as the neutral gas.
\label{fig:MeanVelocity}}
\end{figure}

Gas in the SMC has a wide velocity distribution. The first moment
of the \hi\ gas ranges from +120 km s$^{-1}$ near the SMC-Stream
interface to +190 km s$^{-1}$ near the Bridge. These trends can be
seen in Figure \ref{fig:MeanVelocity}.

Previous studies have shown that the \hi\ component has a strong gradient,
 suggesting that the large scale motions of the \hi\ gas is dominated by rotation 
 \citep{stanimirovic2004anew,DiTeodoro2019}.  In contrast, the \ha\ 
rotation is less pronounced. We picked two pointings along
the greatest velocity gradient line as measured in \citet{stanimirovic2004anew}. Comparing mean velocities
from $\lb = (299\fdg3 ,-45\fdg2)$ and $\lb = (306\fdg0 ,-43\fdg0)$, we find a 
velocity gradient of $+67$ \kms\ for the neutral gas and $+15 $ \kms\ for the ionized gas.
The appearance of weaker rotation is
partially due to the broader nature of the \ha\  lines. WHAM measures
the total integrated emission along the line of sight, and cannot
distinguish a singular cloud with a broad velocity range from several
clouds in the halo of the galaxy imposed over clouds embedded in the
interior or multiple halo clouds at different velocities. Additionally, with our 1\dg\ beam, the small
scale structure is blended together.
 This convolution  broaden the lines, shifting the mean velocity.

\subsection{$H^0$ and $H^+$  Mass \label{sub:Mass}}

We calculated the mass of ionizing gas following \citet{hill2009ionized}
and \citetalias{barger2013warmionized}. We assume a uniform density distribution to
estimate a characteristic line of sight depth:

\begin{equation}
EM=2.75\left(\frac{T}{10^{4}\,\textrm{K}}\right)^{0.924}
\left(\frac{\iha}{\textrm{R}}\right)
\textrm{cm} ^{-6}\,\textrm{pc}  \label{eq:EMEQ}
\end{equation}
The resulting mass in each beam is then: 
\begin{equation}
\left(\frac{M_{\textrm{H}^{+}}}{\msun}\right) = 3.4\times10^{4}\,\Omega 
\left(\frac{D}{\text{kpc}}\right)^{2}\left(\frac{EM}{\textrm{cm}^{-6} \textrm{pc}}\right)\left(\frac{n_{e}}{\text{cm}^{-3}}\right)^{-1},
\end{equation}
where $n_e$ is the electron density, which we estimate from several scenarios described below, and $\Omega$ is the solid angle over which the mass is calculated.
In our case, the solid angle is $1\arcdeg$ WHAM beam or a $0\fdg25$ pixel
in the resampled images. $M_{\textrm{H}^+}$ is the total gas
 mass where H is $\textrm{H}^+$ since we include a
 factor of 1.4 to account for helium as in \citet{hill2009ionized} 
 and \citetalias{barger2013warmionized}. The distance to any section
of the SMC is taken as the average distance of the SMC, or 60 kpc
\citep{Hilditch2005}, with the exception of the SMC Tail region marked
in Figure \ref{fig:Skematic}, where the mass is calculated with
a distance of both 55 kpc and 60 kpc. All mass calculations were determined
by integrating the emission over the $+90 \leq \vlsr \leq +210$ \kms\
velocity range.
We use the methods described in \citetalias{barger2013warmionized} 
to calculate the mass of the \hi\ and include a factor of 1.4
to account for helium. In all cases,
we smooth the \hi\ emission to match WHAM's resolution,
 integrating over the same velocity range as the \ha\ emission.

\begin{table*}[tb]

\caption{Neutral and Ionized Properties\label{tab:ModelTable} }
\begin{center}
\begin{tabular}{cccccccc}

\hline 
\hline 
{Region} & & \multicolumn{4}{c}{Neutral Properties} &  &\tabularnewline
& \cline{2-6} 
 & & $ \log \left\langle \nhi \right\rangle $ & $M_{\mathrm{H}^{0}\mathrm{avg}}$\tablenotemark{a} & $M_{\mathrm{H}^{0} \mathrm{beam}}$\tablenotemark{b} & $L_{\mathrm{H}^{+}}$ & $\log \left\langle n_0 \right\rangle $ & \tabularnewline
 & & [cm$^{-2}]$ & [10$^{6}$ \msun] & [10$^{6}$ \msun] &[kpc] & [cm$^{-3}$] &\tabularnewline
\hline 
SMC\tablenotemark{d} & & 21.0 & 841\tablenotemark{c}& 683\tablenotemark{c} & 9.0 & -1.4 & \tabularnewline
\hi\ SMC Tail\tablenotemark{d,g} & & 20.5 & 160 -- 243 & 166 -- 198& 1.9 & -1.3 &  \tabularnewline
\ha\  SMC Tail\tablenotemark{d,g} & & 20.7 & 112 -- 170& 114 -- 136&1.9 & -1.0 & \tabularnewline
SMC Filament\tablenotemark{d} & & 19.7 & 42& 34&3.3 & -2.3 &  \tabularnewline
\hline 
\hline 
{Region} & &  \multicolumn{4}{c}{Ionized Properties} &  \tabularnewline
& &   \multicolumn{2}{c}{Ionized Skin} &   \multicolumn{2}{c}{Cylinder} & \multicolumn{2}{c}{Ellipsoid}   \tabularnewline
\cline{2-3} \cline{4-5} \cline{6-8} 
 & $\left\langle EM\right\rangle $ &  $L_{\mathrm{H}^{+}}$ & $M_{\mathrm{H}^{+}}$ &   $L_{\mathrm{H}^{+}}$ & $M_{\mathrm{H}^{+}}$ & $L_{\mathrm{H}^{+}}$ & $M_{\mathrm{H}^{+}}$  \tabularnewline
 & [10$^{-3}$ cm$^{-6}$ pc] &  [kpc] & [$10^{6}$ \msun] &  [kpc] & [$10^{6}$ \msun] &  [kpc] & [$10^{6}$ \msun] \tabularnewline
\hline 
SMC\tablenotemark{d}& 8488 & 6.6 & 738\tablenotemark{e} &  9.0 & 788\tablenotemark{e} &14 -- 20\tablenotemark{f} & 671 -- 802\tablenotemark{e} \tabularnewline
\hi\ SMC-Tail\tablenotemark{d,g} & 771 & 0.2 -- 0.3 & 26 -- 34  &    1.9 -- 2.0 & 74 -- 91 &1.9 & 76 -- 90  \tabularnewline
\ha\  SMC-Tail\tablenotemark{d,g} & 1284  & 0.1 -- 0.1 & 11 -- 14 &    1.9 -- 2.0 & 39 -- 48 & 1.9 & 42 -- 50 \tabularnewline
SMC-Filament\tablenotemark{a,h}  &698 & \nodata & \nodata  & 3.3 & 128 & 3.3 & 61   \tabularnewline
\hline 
\end{tabular}
\end{center}

\tablenotetext{a}{Mass range calculated using  column density average.  Applied to ionized skin and cylinder scenarios.} 

\tablenotetext{b}{Mass range calculated using per beam column densities. Applied only to ellipsoid scenario}

\tablenotetext{c}{In all cases, our masses calculated reduces the resolution of the \hi\ to match the WHAM resolutions. This results in an overestimation of the 
\hi\ mass and our results do not match previously published results in \citep{bruens2005theparkes}.} 

\tablenotetext{d}{Regions defined by an ellipsoid (SMC) centered at $\ell=301\fdg6$
and $b=-44\fdg8$ and polygons with
the following corners: $\ell=(301\fdg8,295\fdg9,289\fdg1,295\fdg7)$
and $b=(-39\fdg4,-36\fdg2,-43\fdg0,-46\fdg5)$
for the \hi\  SMC-Tail, $\ell=(300\fdg5,294\fdg5,292\fdg0,297\fdg0)$
and $b=(-41\fdg0,-39\fdg5,-42\fdg0,-45\fdg0)$
for the \ha\  SMC-Tail, and $\ell=(309\fdg0,299\fdg0,299\fdg0,309\fdg0)$
and $b=(-48\fdg0,-48\fdg0,-55\fdg0,-55\fdg0)$
for the SMC-Filament.} 

\tablenotetext{e}{Not including the internal extinction of the SMC results in a mass range of 692 $\times\,10^{6}$ \msun for the ionized skin scenario, 763 $\times\,10^{6}$ \msun for the cylindrical scenario, and  571--698 $\times\,10^{6}$ \msun
for the ellipsoidal scenario.}

\tablenotetext{f}{Line of sight depth at the maximum depth. We use 14 kpc from \citet{Subramaniam2012}, 20 kpc from \citet{Scowcroft2016}. Minimum depth of 3 kpc.}

\tablenotetext{g}{Two different distances are used, 55 kpc and 60 kpc. The average distance
to the Bridge is 55 kpc, however the SMC is measured at 60 kpc. Both
distances are used to provide a range of possible masses and to provide a more direct comparison to results from \citetalias{barger2013warmionized}.} 

\tablenotetext{h}{Ionized skin method not applied here due to little to no \hi\
 column density.} 

\end{table*}

\subsection{Mass of Ionized Gas\label{sub:Mass-of-Ionized}}

To determine the ionized mass of the SMC and its extended system,
we must make assumptions for either the line of sight depths, or the
electron density. While there are line of sight depths in the literature
\citep{Subramaniam2009,Subramaniam2012,Scowcroft2016,ripepi2017thevmc},
these depths are for the stellar component of the SMC. As is apparent
from the extent of the gas, and from Milky Way studies of the WIM,
the \ha\  gas likely has a larger line of sight depth than the stellar
line of sight. For our mass estimates, we assume three different scenarios.
The first scenario assumes the ionized gas is in a skin
around the neutral gas.  For the second and third scenarios,
we assume the gas is well-mixed and the \hi\  line of sight
is equal to the \ha\ line of sight. These line of sight depths must be assumed, as there
is no comprehensive model of \hi\ or \ha\  line of sight
depths.  Our second and third scenarios use a \text{``cylindrical"} and an ``ellipsoidal" geometry, respectively. 
We define the third scenario as our ellipsoid scenario, 
where we assume a standard shape and create a 
model of the structure of the galaxy. For
the mass calculations, we partitioned the galaxy into four different
regions: the SMC Tail, the central galaxy, the SMC-Stream interface,
and the extended ionized gas (Figure \ref{fig:Skematic}).  The SMC Tail 
 is divided into two regions in order to directly compare to
deeper observations made of the same regions in \citetalias{barger2013warmionized}. Values for each
region can be found in Table \ref{tab:ModelTable}. $M_\mathrm{ionized}/M_\mathrm{ionized+neutral}$
values are listed in Table~\ref{tab:IonizationTable}.

In the first ionized skin scenario, we assume that the ionized gas lies in a skin around the \hi\ features.
This assumes a relation between the neutral gas density and the density of the electrons.
This results in two possibilities, $n_{e}=n_{0}$ and $n_{e}=n_{0}/2$. If the neutral
and ionized components are separated, but are in pressure equilibrium,
then the electron density of an ionized skin would equal half the neutral
hydrogen density \citep{hill2009ionized}. However, because the recombination
time ($\sim 1$ Myr) is much shorter than the sound
crossing time (a few hundred Myr), we expect that $n_{e}\approx n_{0}$
is more likely and only consider this case.
We assume the \hi\ column densities
from HI4PI to calculate the electron column densities.
This method is more reasonable outside the center
of the galaxy, in regions where there is significant \hi\
emission. Since we use the average column density 
over each region to calculate the mass, we use the averaged extinction over the same region to be consistent.
The ionized hydrogen mass estimate for the SMC using this first method
 is then $\sim7.4\times10^{8}\,\msun$.

Our second scenario assumes a cylindrical geometry. In this scenario, 
we assume $L_{\mathrm{H}^+} = L_{\mathrm{H}^0}$.
This treats the ionized and neutral gas as components occupying similar mixed
volumes/regions along the line of sight. In this case, the line of sight
of the neutral component is similar to the line of sight of the ionized
component, and each region is assumed to have the same path length, effectively 
a cylindrical geometry. The intensity is then averaged over the region, and treated
as if it is in a single flat layer, with a depth equivalent to the width of the semi-major axis of the \hi\ feature.
 This scenario uses the average extinction for the region. This results in a mass estimate
of $\sim7.9\times10^{8}\,\msun$.

The third scenario is our ellipsoid scenario. Similar to our cylindrical scenario,
 we assume that neutral and ionized gas lies along the same line of sight. However,
we model the shape of the SMC as a simple
ellipsoid, in order to estimate the line of sight depth of the
gas. Similar to the \citetalias{barger2013warmionized} method for the Bridge, we assume that the maximum
depth is equivalent to the width of the galaxy. We define the width
as the semi-major axis. We chose two points on
the edge of the $\nhi = 1.0\times10^{20}\,\mathrm{cm}^{-2}$
contour to define the semi-major axis, centered at the radio kinematic
center as defined in \citet*{stanimirovic2004anew}. The projected
ellipse can be seen in Figure \ref{fig:Skematic}, with a semi-major
axis of 6 kpc and a semi-minor axis of 4.55 kpc. We then assume that
the line of sight depth of the ellipse is from 14 kpc \citep{Subramaniam2012}
to 20 kpc \citep{Scowcroft2016}, based on measurements from red clump
stars, RR Lyrae stars, and Cepheids. We constrained the minimum depth
for the model to be no less than 1.9 kpc, again using the assumption
that the width of the extended areas are similar to the depths. While
there are estimates of the SMC's inclination, we chose to leave more
extensive models for future work. Sightlines with intensities below
0.01 R were excluded in the mass calculations. In this scenario, we calculate the extinction
correction and mass within individual smoothed 0\fdg25 
pixels to stay consistent with the mass calculation methods. Using this model, we
found the ionized content of the SMC to be $\sim6.7\times10^{8}\,\msun$
using a depth of 14 kpc and $\sim8.0\times10^{8}\,\msun$
using a depth of 20 kpc. 

\begin{table*}[tb]
%\centering{}
\caption{Ionization Mass Fraction: $M_\mathrm{ionized}/M_\mathrm{ionized+neutral}$\label{tab:IonizationTable} }

\begin{center}
\begin{tabular}{cccc}
%\multicolumn{4}{c}{Ionization Mass Fraction: $M_\mathrm{ionized}/M_\mathrm{ionized+neutral}}}\tabularnewline
\hline 
\hline 
 & Ionized Skin ($n_e = n_0$) & Cylinder& Ellipsoid \tabularnewline
\hline 
SMC & 46\% & 48\%& 50\% -- 54\%  \tabularnewline
\hi\ SMC-Tail & 12\% -- 14\% & 27\% -- 32\%& 29\% -- 31\% \tabularnewline
\ha\  SMC-Tail & 8\% -- 9\% & 22\% -- 26\%& 24\% -- 27\% \tabularnewline
SMC-Filament\tablenotemark{a} & \nodata & 79\%& 65\% \tabularnewline
\hline 
\end{tabular}
\end{center}

\tablenotetext{a}{Ionized mass fraction not calculated using $n_e = n_0$ due to little to no \hi\ column density}

\end{table*}

These simplified scenarios allow us to make estimates of the extended
ionized gas mass but do not include many effects that would be
 important for higher-resolution data. The gas is more likely
to be clumpy and discontinuous, with varying densities and components
at different distances. These models also do not consider 
 the inclination of the SMC nor how that would change the projected
line of sight. Our calculations assumed an average distance of 60
kpc. However, gas toward the Stream is more distant,
and gas toward the Bridge is closer to 55 kpc. In all regions, the \hi\ resolution is reduced to
 match the WHAM resolution to consistently compare
emission along the same line of sight. The results of this approximation, along with a different 
defined area for the galaxy, is a higher calculated \hi\ mass than previously measured in
\citet{bruens2005theparkes}. All three scenarios give $M_\mathrm{H^+} \sim 6\times10^8 \msun$.
We do not quantify the uncertainties in the many assumptions involved, but the consistency of the
 estimates gives us confidence that they are realistic.

\citetalias{barger2013warmionized}  uses neutral column densities from the LAB
\hi\ survey \citep{kalberla2005theleidenargentinebonn} data
set for the mass calculations of the region; here we use HI4PI. The
use of HI4PI results in $\sim4$--$7\%$ increase in \hi\ in
the \ha\  and \hi\ Tail, which in turn decreases the mass
\ha\  calculated for the assumed case in which we set the density
of the electrons, $n_{e}$, equal to that of the neutral gas $n_{0}$.
 By also included a 20\% correction which
 accounts for degradation in the WHAM instrument and using HI4PI instead of LAB,
we find similar mass results for the velocity range $+100  \leq$  \vlsr\ $ \leq +210$ \kms.

\section{Discussion \label{section:Discussion}}

\subsection{SMC Filament}

The trailing Magellanic Stream (MS) was first discovered by 
 \citet{Mathewson1974}. \citet{Putman2003}  provided higher-resolution
  observations that strongly suggest a bifurcated, filamentary structure. Through 
 a Gaussian decomposition of the \hi\ emission, \citet{nidever2008theorigin}
  traced these two filaments to the LMC and the Magellanic Bridge. 
  The ``LMC" filament is marked in Figure \ref{fig:Comparison-of-HI} with a dashed red line.  However, the right filament in solid red has a less
certain origin. \citet{Fox2010} found that quasar sightlines
within the shorter filament have abundances of roughly 0.1 solar metallicity.
Such measurements match the expected metallically of the SMC 2.5 Gyrs
ago \citep{pagel1998chemical,harris2004thestar}, further suggesting
an SMC origin for the filament. 

The origin of the filament appears to be the lower left side of the
SMC, toward the Stream, which is marked in Figure \ref{fig:Comparison-of-HI}
(bottom right). This could suggest that the \ha\  filament is a shock,
similar to shocks presented in \citet{bland-hawthorn2007thesource}.
The \ha\  filament is near the SMC \hi\ filament and could
be the leading edge of the neutral SMC filament (solid red), pulled
out similarly to the Stream filaments, and ionized by shocks. In \citet{Mastropietra2005}
simulations were run with the LMC and MW, where even a low-density
halo removed significant gas from the system. While the simulations
did not include the SMC, if similar conditions strip gas out of the
SMC, proper motions from \citet{Killivayalil2013} suggest gas stripped
would trail the SMC in the direction of the Stream. Overtime, this
filament may have become more highly ionized, leaving only a few scattered
\hi\ clouds intact. Or, if the SMC has a puffy, ionized halo,
interactions could be stripping outer gas directly. Confirming the
extent of the \ha\  beyond the $b=55\arcdeg$ edge of our survey
will help give a clearer picture, and line ratios could also help
point to the origins of the \ha\  enhancement. A recent study by \citet{McClure-Griffiths2018} found
a neutral outflow centered at $(\alpha,\delta) \approx 00^{h}49^{m}, -71\dg 15'$, near the origin of the ionized filament. While the extent of the ionized filament is longer
then the neutral outflow and offset spatially, exploring a possible relationship between the two features could help 
identify the origins of the structure.
Extended observations of the filament is part of the ongoing WHAM Magellanic Stream Survey
in \ha.\

 \subsection{Extended SMC Gas}

The diffuse ionized gas of the SMC extends beyond
   the traditionally defined extent of the neutral gas. The morphology, 
   kinematics, and environment
   of this gas have been explored in several studies.
  
  The COS/UVES survey compared five sightlines along the Magellanic system with WHAM
   observations \citep{fox2014thecosuves}. Along one sightline, FAIRALL9, they found an offset in 
   velocity between the neutral emission at \vlsr\ = +190 \kms and the \ha\ emission at \vlsr
    = +156 \kms, with the \ha\ offset confirmed by
   the detection of \cii\ at \vlsr\ = +150 \kms. While the location of FAIRALL9 is just outside the SMC map, 
   we find a similar offset along a significant number of our observations for the SMC and the 
   surrounding material, which can be seen in Figure \ref{fig:MeanVelocity}. 
  They suggest this could indicate that ionized gas exists as a boundary layer that is compressed 
  by ram pressure as the neutral gas moves through an ionized medium.
\citet{Diaz2011} modeled the Magellanic System and found the bifurcation of the MS was sensitive to
 drag from the hot halo, limiting the velocity dispersion of the MS structures. While this model is not able 
 to reproduce the detailed kinematic and spatial structure of the Magellanic Stream,
  it does suggest that interactions with the surrounding coronal gas affect the morphology of the Stream. Similarly, in \citet{barger2017}, an analysis of a sample of  \ha\
  observations throughout the Stream suggests that halo-gas interactions
   likely affect morphology even if they are not the primary source of ionization.
    
  The effect of the MW corona on the MC's is also discussed
   in \citet{Salem2015}. They find the interaction
  between the MW coronal gas and the LMC results in a sharp 
  drop-off in the radial gas profile. While they are not able to similarly constrain
   the SMC, they find that interactions with the corona would 
  result in a truncated halo $\approx$ 3.5 kpc in size. Their study only 
  considered the neutral gas in the SMC with a correction for the ionized gas mass, but we find that the denser
neutral gas ends more closely to the main body of the galaxy and 
does not extend out as far as the diffuse ionized gas. This picture is\
more consistent with \citet{Wang2019} as their models trace
the morphology of the warm and hot gas independently from
the neutral component. Modeling the formation of the MS 
and including effects of ram pressure stripping on the Clouds,
 they find a large, extended envelope of ionized gas
while the neutral gas is concentrated near the galaxies.

  Pairing the WHAM observations of the extended SMC with further
  absorption studies and dynamical models may help to
   constrain and explore the effects of the MW corona and ram pressure stripping on the system.

\subsection{External Galaxies}
We have measured diffuse emission in the halo of the SMC with sensitivity an order 
of magnitude better than other studies that have looked at the DIG in external galaxies. 
Observations from \citet{Rossa2000} and \citet{Dettmar1990} 
use a similar method to estimate the total DIG mass of edge on galaxies. 
In both studies,  a filling factor is assumed and applied to the 
emission measure and a line of sight distance is assumed  with a correction factor included for helium. The resulting
 masses are on the order of $10^{6} - 10^{8}\ \msun$.
 For the galaxy with the largest observed diffuse gas mass, NGC 891, the 
 total calculated diffuse gas mass is $4.0\times10^{8}\ \msun$ \citep{Dettmar1990}. 
 This can be compared to its neutral halo mass of $\rm{M}_{H^0} \approx 1.2\times10^{9}\ \msun$  \citep{Oosterloo2007}. 
 In this case, the measured mass contribution of the low-ionization 
 species is lower than the comparable neutral
 component. However, in our study, we are able to look through the galaxy and measure both diffuse gas
 in the halo as well as in the disk. In addition to the difference in sightlines, \citet{Rossa2000} is able to detect 
the DIG down to a few Rayleighs, compared with WHAM's $\sim25$ mR sensitivity for 
singular pointings and $\sim10$ mR for continuous emission. To directly compare 
our observations to the detection limits of the previous studies,
we can limit our mass calculations to bright emission. When only measuring the mass of
 regions with emission 5.0 R and over, which coincides with regions in the SMC
  with \nhi\ $> 5.0 \times 10^{20}$ and is comparable to the extent of MCELS, we see the ionized mass of the SMC reduced by $\sim50\%$. 
  
Several other studies have looked at the fraction of 
\ha\ emission originating from the WIM compared to \hii\ regions \citep{Hoopes1996,Ferguson1996,Wang1999,Oey2007}. These studies find the total contribution of \ha\ emission from the WIM ranges from $30\%$ to $59\%$. If we use this guidance to estimate the fraction of observed emission arising from WIM within the core of the galaxy, the mass of ionized gas in this region is reduced by about $25-45\%$. While we cannot separate these contributions easily due to our large beam size, such estimates still result in a substantial WIM mass for the SMC, even within its traditional boundaries.

\subsection{Future Work}

In this work, we present the first detection of the extended ionized halo of the
 SMC and the first measurements of its associated mass. While some 
 spectra show evidence for emission outside the range we focus on
  ($+90 \leq \vlsr \leq +210\, \kms$), confusion with MW  \ha\ emission and OH 
  skyline emission prevents us from including these regions using this first dataset.
   \citet{nidever2008theorigin} use Gaussian decomposition to trace extended \hi\ structures 
   carefully through velocity regions where the Magellanic Stream blends with other sources. 
   Combining our knowledge of MW emission from the WHAM Sky Survey \citep{haffner2003thewisconsin}
   and these methods may help to isolate SMC emission in the future.

 While Gaussian decomposition may help with emission at lower velocities, we are limited
  at higher velocities due to our window only including the edge of the OH lines. 
  \citetalias{barger2013warmionized}
  was able to recover some Magellanic emission (with reduced sensitivity) in 
  spectra where the OH line was more completely sampled within their velocity 
  window. We would need to obtain additional observations to attempt similar 
  extractions in the areas of this study that are affected.

\section{Summary \label{section:Conc}}

We have mapped the extended \ha\  emission of the SMC using WHAM covering
a total of 546 square degrees over $+90 \leq \vlsr \leq +210\,$ \kms. We compare these observations to the 21 cm emission from the
HI4PI \hi\ survey. Through these observations, we examine
the extent, morphology, velocity gradients, and mass of the ionized
gas. The main conclusions from our work are: 

\begin{enumerate}
\item \textbf{\nhi\ and \iha\ distributions}: In the center of the SMC, there is a strong correlation
between the \nhi\  and \iha. The similar structure holds for column densities \nhi $> 10^{20}$ cm$^{-2}$. 
Beyond the dense inner region, the \ha\ emission extends further with intensities of 0.03-0.18 R. At these intensities, there is
less direct correlation with the \nhi. This suggests
there are highly ionized regions where little neutral gas remains.
Of particular note is the \ha\ filament originating at $\lb = (305\fdg5, -50\fdg0)$ and extending below the galaxy. With only scattered
corresponding \hi\ clouds, this region appears to be highly
ionized. This survey is limited to velocities between $+90 \leq \vlsr \leq  +210$ \kms\ due to atmospheric OH and Milky Way emission contamination.

\item \textbf{Velocity distribution}: The rotation of the gas in the galaxy
seen in \nhi\ and \iha\ velocity
maps match well (Figure \ref{fig:MeanVelocity}). This suggests the ionized
gas is kinematically related to the neutral component. Although the
global velocity trends agree, the intensity-weighted average \ha\ 
velocities are offset compared to the neutral gas, with the offset
dependent on location. Multi-component distributions may contribute
to the offset in velocity. 

\item \textbf{Ionized gas mass}: If we assume a distance of 60 kpc, with
our simplified models we find an ionized gas mass of the central SMC
to be 6.7 -- $8.0 \times 10^{8}\ \msun$. With a central neutral mass
of $6.8 - 8.4 \times 10^{8}\ \msun$,
the ionized gas appears to be roughly one half of the total atomic
gas mass. The total ionized mass of all regions is $(0.8-1.0)\times10^{9}\,\msun$,
which is comparable to the total neutral mass in the same region of
$(0.9-1.1)\times10^{9}\,\msun$. Assumptions for $n_{e}$ and line of-sight distances
dominate our uncertainty in the mass calculations. 

\end{enumerate}

\section{Acknowledgements}

We acknowledge the support of the U.S. National Science Foundation (NSF) for WHAM development, operations, and science activities. We thank Bob Benjamin for his comments on the manuscript. The survey observations and work presented here were funded by NSF awards AST 1108911 and 1714472/1715623. We also thank the staff at CTIO for their continued support. We thank the referee for constructive comments that have improved the discussion in this paper.

\bibliography{smcbib}
\bibliographystyle{aasjournal}
\end{document}